\documentclass[pdflatex,sn-mathphys,referee]{sn-jnl}% Math and Physical Sciences Reference Style

\usepackage{amsmath}
\usepackage{siunitx}
\usepackage{xr}
\usepackage{comment}
\usepackage{lineno}
\usepackage{caption}

\usepackage{bm}
\usepackage[labelfont=bf]{caption}
\definecolor{Highlight}{HTML}{bae0a4}

\makeatletter

\newcommand*{\addFileDependency}[1]{% argument=file name and extension
\typeout{(#1)}% latexmk will find this if $recorder=0
\@addtofilelist{#1}
\IfFileExists{#1}{}{\typeout{No file #1.}}
}\makeatother

\newcommand*{\myexternaldocument}[1]{%
\externaldocument{#1}%
\addFileDependency{#1.tex}%
\addFileDependency{#1.aux}%
}
\myexternaldocument{Supporting_Info}

\theoremstyle{thmstyleone}%
\theoremstyle{thmstyletwo}%

\theoremstyle{thmstylethree}%

\begin{document}

\title[Intrinsic plasmon canalization in the biaxial van der Waals crystal  $\mathrm{MoOCl}_{2}$]{Intrinsic plasmon canalization in the biaxial van der Waals crystal $\mathrm{MoOCl}_{2}$}

\author[1]{\fnm{Farid} \sur{Aghashirinov}}
\equalcont{These authors contributed equally to this work.}

\author*[2]{\fnm{Andrea} \sur{Mancini}}\email{andrea.mancini@iit.it}
\equalcont{These authors contributed equally to this work.}

\author[2]{\fnm{Lin} \sur{Nan}}

\author[2]{\fnm{Giacomo} \sur{Venturi}}

\author[1]{\fnm{Bettina} \sur{Frank}}

\author*[1]{\fnm{Harald} \sur{Giessen}}\email{giessen@pi4.uni-stuttgart.de}

\author*[2]{\fnm{Antonio} \sur{Ambrosio}}\email{antonio.ambrosio@iit.it}

\affil[1]{\orgdiv{4th Physics Institute and Research Center SCoPE}, \orgname{University of Stuttgart}, \orgaddress{\city{Stuttgart}, \postcode{70569}, \country{Germany}}}

\affil[2]{\orgdiv{Centre for Nano Science and Technology}, \orgname{Italian Institute of Technology}, \orgaddress{ \city{Milan}, \postcode{20134}, \country{Italy}}}

%TC:ignore

\abstract{Anisotropic polaritons in low-symmetry crystals allow for subwavelength confinement and directional routing of light. The most extreme form of such anisotropy arises at the topological transition between elliptical and hyperbolic dispersion, where the isofrequency contours collapse into parallel lines and polaritons propagate in a diffractionless, beam-like fashion. This canalization regime has previously been accessed through twisted heterostructures or engineered metasurfaces. Here we show that natural canalization can be achieved without any fabrication or structuring by exploiting the intrinsic elliptical-to-hyperbolic transition in the van der Waals crystal $\text{MoOCl}_{\textnormal{2}}$ at room temperature. Using near-field imaging, we directly visualize plasmon-polariton canalization emerging at the low-loss Drude crossing point along the [010] crystal axis. Owing to the moderate slope of the Drude permittivity, the resulting polaritons remain highly directional across a broad spectral window. This weak dispersion also enables robust thickness-dependent tuning, and we demonstrate, both experimentally and theoretically, that the canalization wavelength can be adjusted by more than \SI{1}{\micro \meter} simply by varying the flake thickness. This work brings canalized polariton propagation into the 4.5 - \SI{6}{\micro \meter} range, beyond the frequency limits of phonon-polariton platforms and overlapping with important molecular vibrations, opening new opportunities for mid-IR nanophotonics and sensing.}

%TC:endignore

\maketitle

\pagestyle{plain}

\section{Introduction}\label{sec1}

Polaritons in van der Waals (vdW) materials provide extreme electromagnetic field confinement \cite{ciraci2012probing, li2015hyperbolic, alcaraz2018probing, herzig2024high, kowalski2025ultraconfined}, strong nonlinear responses \cite{Zhang2021, datta2022highly}, and widely tunable dispersion \cite{chen2012optical, fei2012gate}, offering a powerful platform for nanoscale control of optical energy \cite{Basov2016, zhang2021interface,low2017polaritons, alvarez2022negative, hu2023gate,sternbach2023negative,teng2024steering, zhang2025phonon, jackering2025tailoring}. In anisotropic crystals, polaritons can exhibit hyperbolic dispersion with open isofrequency contours (IFCs), enabling highly directional and deeply sub-diffractional propagation \cite{hu2020phonon, he2022anisotropy, passler2022hyperbolic, zhou2026fundamental}. Such behavior has been extensively explored in vdW crystals supporting phonon polaritons, where lattice vibrations couple to electromagnetic fields to form strongly confined modes in the mid-infrared (IR) \cite{Dai2014, Ma2018, zheng2019mid}. Recently, low-loss natural hyperbolicity has been extended to visible and near-IR frequencies via real-space imaging of plasmon–polariton propagation in the vdW crystal MoOCl$_2$ \cite{Venturi2024,ruta2025good,li2025broadband, zhang2025manipulating, ghosh2026spatiotemporal}, enabling access to natural hyperbolic polaritons in a spectral range that was previously inaccessible.

Among anisotropic dispersion phenomena, canalization is of particular interest as it occurs at the topological transition between hyperbolic and elliptical dispersions, where polaritons propagate without diffraction, forming highly directional, beam-like modes. Canalization has been realized through engineered in-plane anisotropy using metasurfaces \cite{gomez2015hyperbolic, Correas-Serrano2017, Li2020, xu2026programmable}, twisted heterostructures \cite{hu2020topological,duan2020twisted,chen2020configurable,obst2023terahertz,duan2023multiple,li2025broadband, Zhou2025}, and substrate engineering \cite{duan2021enabling, Hu2022, duan2025canalization,zhu2025multiple,zhang2025ultimate, chang2022field}. A conceptually simpler approach to achieve canalization is to exploit the natural hyperbolic to elliptical transition occurring when one component of the permittivity changes sign. However, despite its simplicity, this effect has thus far been mainly explored in phonon-polariton platforms, where its observation is often hindered by the spectral structure of the dielectric response. In many cases, the transition frequency lies close to a phonon resonance, leading to increased damping and limiting the visibility of propagating canalized modes \cite{Terán-García, Wang, TresguerresMata2024, Díaz-Núñez,zheng2024hyperbolic, shiravi2026tunable}, in addition to the intrinsically narrow bandwidth of Reststrahlen bands and, in some systems, the need for out-of-plane response \cite{ou2025natural} or low-temperature operation \cite{Liu2025}. To date, the only reported exception is the broadband plasmon canalization found in 2M-WS$_2$, where the reduced steepness of its Drude dispersion enables canalized propagation over an extended wavelength range \cite{Xing2024}. This observation points to a general materials strategy for broadband canalization based on plasmonic anisotropy rather than phononic resonances. However, proof of intrinsic canalization in 2M-WS$_2$ relied on patterned structures to indirectly probe the polaritonic IFCs, precluding a direct, real-space visualization of canalized polariton propagation \cite{Xing2024}.

Here, we demonstrate broadband plasmon-polariton canalization in MoOCl$_2$ at mid-IR frequencies and at room temperature through direct near-field real-space mapping. We visualize the transition from hyperbolic to elliptical propagation across the 2.3–7~\textmu m wavelength range and corroborate our observations with numerical simulations and analytical calculations of the isofrequency contours. Furthermore, we theoretically predict that the canalization wavelength can be tuned over more than \SI{1}{\micro\meter} by varying the flake thickness. This trend is experimentally confirmed through Fourier analysis of the near-field patterns, where the vanishing of the IFC curvature serves as a signature of the topological transition. In contrast to conventional phonon-polaritonic platforms governed by Lorentz-type dispersions, the Drude-like response of MoOCl$_2$ enables broadband canalization over an extended spectral range while simultaneously providing efficient thickness-dependent tunability. The observed shift of the canalization wavelength is primarily driven by the dispersion along the [010] axis, whereas the strongly metallic response along [100] remains deeply negative and therefore only weakly influences the evolution of the IFC curvature with thickness.

\begin{figure}[h!]
\captionsetup{skip=12pt}
    \centering
    \includegraphics[width=1.0\linewidth]{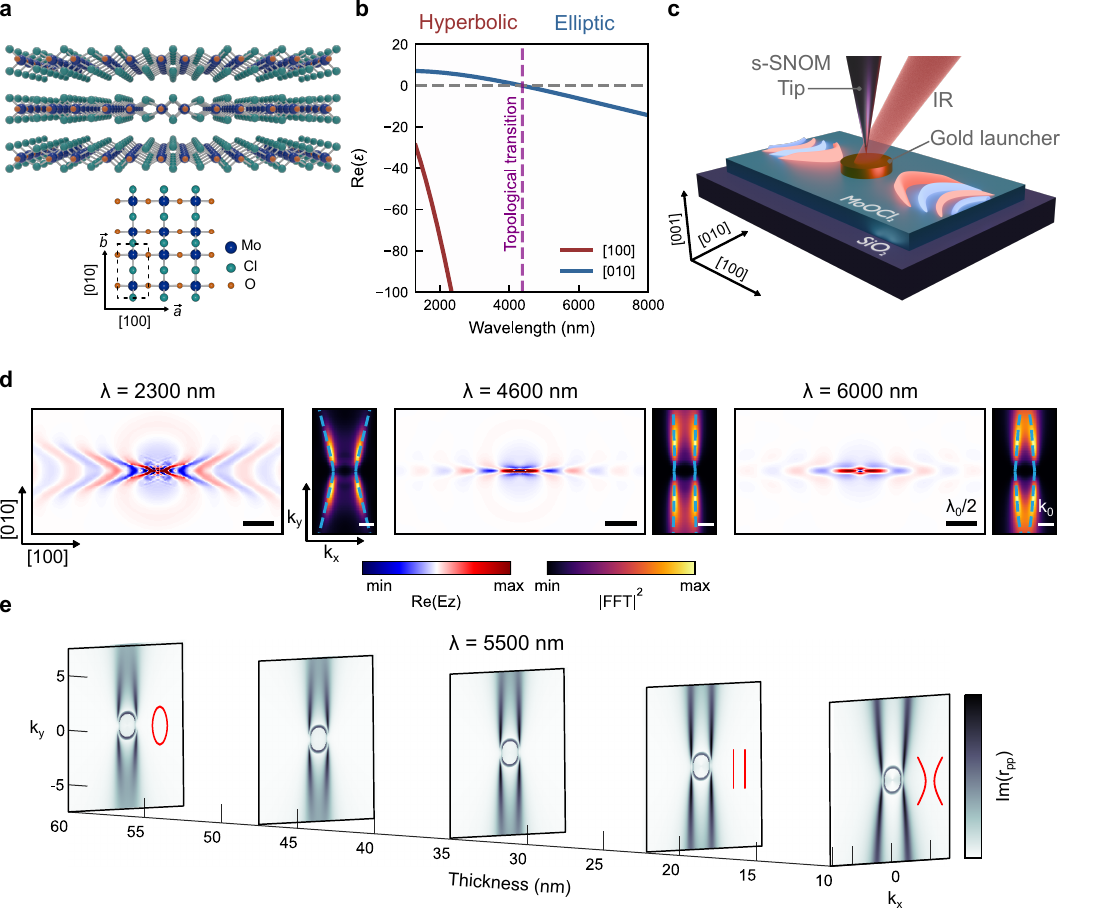}
    \caption{ \textbf{Intrinsic mid-infrared plasmon polariton canalization in hyperbolic $\bm{\mathrm{MoOCl}_{2}}$}. \textbf{a} Crystal structure of bulk $\mathrm{MoOCl}_{2}$ and its in-plane view with atomic 2D unit cell illustrated as dashed lines. \textbf{b} Real part of the dielectric function of $\mathrm{MoOCl}_{2}$ along the $x$ [100] and $y$ [010] directions, showing the transition from the hyperbolic to elliptical regime. The point where $\mathrm{Re}(\varepsilon) = 0$ corresponds to the topological transition frequency, at which canalization is expected. \textbf{c} Schematic of the sSNOM measurement on a $\mathrm{MoOCl}_{2}$ flake where a gold disk nanoantenna is used to launch polaritons. \textbf{d} Simulated near-field out of plane electric field maps of a dipole source placed on top of a $\mathrm{MoOCl}_{2}$ flake with a thickness of 45 nm and corresponding k-space images. Dashed lines are the analytically calculated in-plane plasmon IFCs. \textbf{e} Transfer-matrix calculations at $\lambda$ = 5500 nm for different $\mathrm{MoOCl}_{2}$ thicknesses.}
    \label{fig:fig1}
\end{figure}

\section{Results}
\subsection{Intrinsic SPPs canalization in $\mathrm{MoOCl}_{2}$}

Materials that intrinsically combine strong in-plane anisotropy with broadband plasmonic response are particularly promising candidates for realizing natural plasmon–polariton canalization. Molybdenum oxychloride ($\mathrm{MoOCl}_{2}$) is a layered van der Waals material with a strongly anisotropic crystal and electronic structure \cite{wang2020fermi, Venturi2024,Melchioni2025,li2025broadband, ermolaev2026giant, melchioni2026anisotropic} (Fig.~\hyperref[fig:fig1]{1a}), exhibiting metallic and dielectric responses along orthogonal in-plane directions from the visible to the mid-IR range. Due to an orbital-selective Peierls distortion, the in-plane permittivity components $\varepsilon_x$ and $\varepsilon_y$ have opposite signs over a broad spectral range (Fig.~\hyperref[fig:fig1]{1b}), naturally supporting hyperbolic plasmon polaritons \cite{zhao2020highly}. The gradual Drude dispersion in $\mathrm{MoOCl}_{2}$ further produces a smooth transition between elliptical and hyperbolic regimes, enabling plasmon-polariton canalization without external symmetry breaking (Fig.~\hyperref[fig:fig1]{1b}). The in-plane permittivity in Fig.~\hyperref[fig:fig1]{1b}, which we use throughout our work, is experimentally extracted by fitting reflection and transmission spectra with a three layer model for various flake thicknesses \cite{Melchioni2025} (see Methods and Supplementary Information \ref{Eps_extraction}).

We first confirm the presence of plasmon canalization around the transition frequency where the [010] permittivity changes sign via full 3D electromagnetic simulations (Fig.~\hyperref[fig:fig1]{1d}). We place a near-field source \SI{50}{\nano \meter} above a \SI{45}{\nano \meter} slab of $\mathrm{MoOCl}_{2}$ and record the out-of-plane electric field \SI{5}{\nano \meter} above the surface. By sweeping the dipole emission wavelength from \SI{2300}{\nano \meter} to \SI{6000}{\nano \meter} we can recognize all three propagation regimes: hyperbolic, canalized and elliptical. By 2D Fourier transform (FFT) of the complex field, we obtain the IFCs shape which confirm the observation of the topological transition. The theoretical dispersion calculated in the small thickness limit \cite{alvarez2019analytical, Venturi2024} (dashed blue lines) shows agreement of the model with the computed IFCs. We also observe that the shape of the IFCs around the topological transition can be adjusted by changing the $\mathrm{MoOCl}_{2}$ thickness, allowing for the engineering of the canalization frequency. We reveal this trend by computing the 2D IFCs via transfer-matrix formalism \cite{passler2017generalized} at a fixed wavelength ($\lambda = \SI{5500}{\nano \meter}$) for thicknesses ranging from \SI{10}{\nano \meter} to \SI{60}{\nano \meter} (Fig.~\hyperref[fig:fig1]{1e}). We observe that thinner flakes shows hyperbolic propagation, while thicker ones behave as elliptic, implying that for some intermediate value the IFCs must transition through a canalization point, which is here found around \SI{22.5}{\nano \meter}. This analysis also explains why the canalization observed in the thin film case shown in Fig.~\ref{fig:fig1}d does not occur exactly at the bulk topological transition identified in Fig.~\ref{fig:fig1}b.

\subsection{Real-space imaging of naturally canalized SPPs in $\mathrm{MoOCl}_{2}$ flakes}

\begin{figure}[h!]
\captionsetup{skip=12pt}
    \centering
    \includegraphics[width=1.0\linewidth]{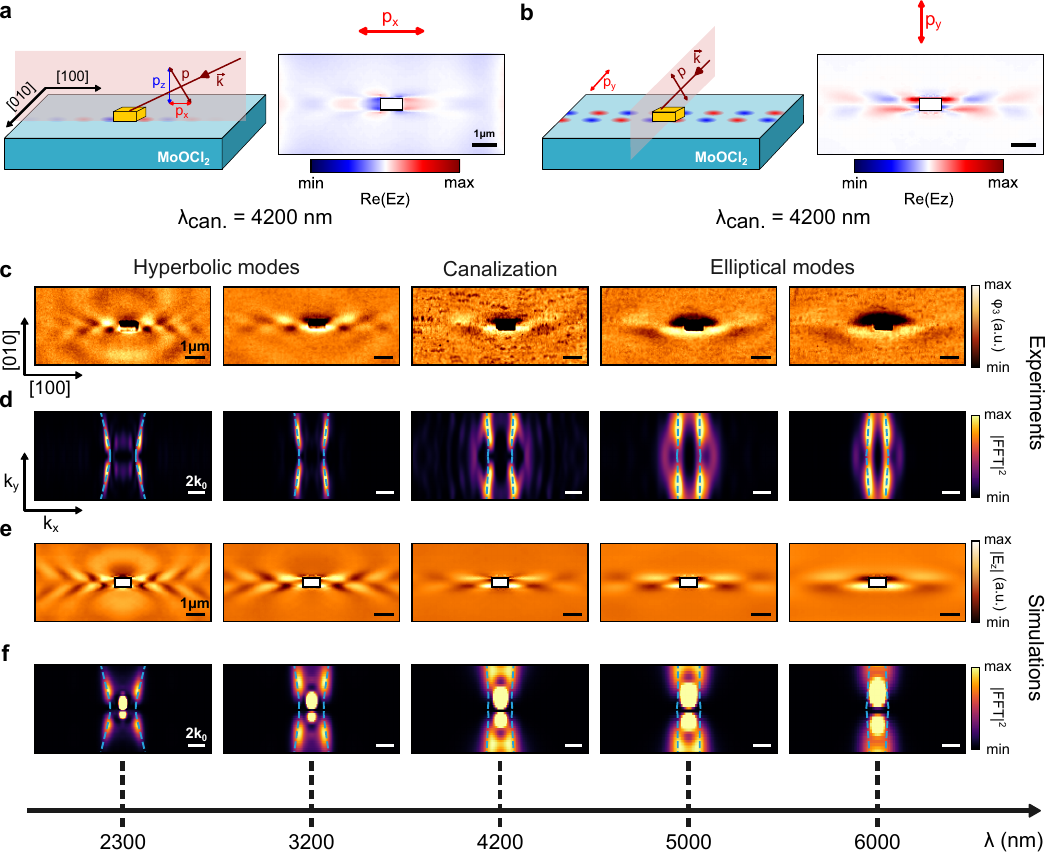}
    \caption{\textbf{Real-space imaging of the intrinsic plasmon polaritons topological transition in $\bm{\mathrm{MoOCl}_{2}}$}. \textbf{a–b} Schematic illustration and simulated near-field maps showing the polarization-dependent launching efficiency of a gold nanorod at the canalization wavelength. The red color arrows illustrate the in-plane $p_{x}$ and $p_{y}$ polarization directions. \textbf{c-d}  Experimental sSNOM phase maps (third demodulation order) of PPs launched by a gold nanorod fabricated on top of the flake (thickness 45 nm) at $\lambda_0$ = 2300, 3200, 4200, 5000, and 6000 nm (scale bars: 1 \textmu m) and corresponding k-space maps with analytical results of the PPs dispersion. \textbf{e–f}~Simulated near-field maps of PPs launched by a gold nanorod fabricated on top of the flake, and the corresponding k-space maps.}
    \label{fig:fig2}
\end{figure}

To experimentally confirm the theoretical predictions presented in Fig.~\hyperref[fig:fig1]{1}, we conduct extensive near-field measurements at various wavelengths on several flakes of different thicknesses. $\mathrm{MoOCl}_{2}$ flakes were obtained by mechanical exfoliation from commercially sourced bulk crystals and subsequently transferred onto 285-nm $\mathrm{SiO}_{2}$/Si substrates (see Methods). Since $\mathrm{MoOCl}_{2}$ is strongly anisotropic, the in-plane polarization plays a crucial role in the excitation efficiency of plasmon polaritons. In the experiments, the incident light is p-polarized to ensure efficient coupling to the s-SNOM tip. When the in-plane electric field component $p_x$ is aligned with the metallic crystallographic axis $[100]$, the launching efficiency of the gold nanoantenna is strongly suppressed, as illustrated in Fig.~\hyperref[fig:fig2]{2a}. Corresponding simulations at $\lambda_{\mathrm{can}} = 4200~\mathrm{nm}$ for a 45~nm-thick flake show rapidly decaying polaritons with only a few visible interference fringes. In contrast, aligning the in-plane polarization $p_{y}$ along the dielectric $[010]$ axis significantly enhances the launching efficiency, resulting in more extended interference patterns (Fig.~\hyperref[fig:fig2]{2b}). For this reason, all subsequent measurements were performed with the in-plane polarization aligned along $[010]$. This behavior can be qualitatively understood in terms of a better overlap between the angular distribution of the in-plane dipole induced in the launcher and the shape of the polariton IFCs at the canalization condition (see Supplementary Information \ref{Polaritons_polarization} for more details and experimental verification).

The visual difference of the field profiles in Fig.~\hyperref[fig:fig2]{2a,b} can be associated with a different instantaneous charge displacement direction with respect to the polaritons wavevector: parallel in Fig.~\hyperref[fig:fig2]{2a} and orthogonal in Fig.~\hyperref[fig:fig2]{2b}. An additional advantage of this arrangement is that no correction of the IFCs obtained from the experimental FFTs is required. This is because the light wavevector is nearly orthogonal to the polariton propagation direction, so the measured fringe periodicity directly corresponds to the polariton wavelength. In contrast, for collinear illumination geometries, the measured periodicity includes an additional contribution from the projection of the free-space wavevector, which depends on the illumination angle \cite{kaltenecker2020mono, mancini2022near}. While this can be advantageous for accessing additional information by simultaneously observing both forward and backward-propagating polaritons with respect to the incident light momentum, it also introduces additional uncertainty in the quantitative extraction of the polariton wavevector due to the angle-dependent momentum contribution.

Near-field phase maps obtained on a 45~nm-thick flake at free-space wavelengths $\lambda_0$ = 2300, 3200, 4200, 5000, and 6000~nm are displayed in Fig.~\hyperref[fig:fig2]{2c}. At $\lambda_0$ = 2300~nm and 3200~nm, hyperbolic wavefronts are observed, originating from plasmon polaritons launched by the gold nanorod, whereas at $\lambda_0$ = 5000~nm and 6000~nm the wavefronts exhibit elliptical shapes. Between the hyperbolic and elliptic regimes, an intermediate propagation regime is expected in which waves propagate without divergence, commonly referred to as the canalization regime. According to Fig.~\hyperref[fig:fig1]{1b}, for a flake thickness of 45~nm, this regime is predicted to occur at a wavelength of approximately 4400~nm. However, this wavelength lies outside the accessible range of our laser source. Consequently, we employed the closest available wavelength that could be generated by our laser system in order to probe the canalization regime. This approach is justified because the dispersion relation shown in Fig.~\hyperref[fig:fig1]{1b} varies slowly in this spectral region, and canalization does not occur at a single, sharply defined wavelength, but rather over a finite wavelength interval around the target value. For these reasons, we use a wavelength of 4200~nm to excite the canalization mode. The transitions between the different propagation regimes can also be clearly observed in momentum space, as shown in Fig.~\hyperref[fig:fig2]{2d}. By applying Fourier-transform analysis to the experimental near-field data, the resulting $k$-space images show open IFCs at $\lambda_0$ = 2300~nm and 3200~nm, corresponding to the hyperbolic regime. At $\lambda_0$ = 4200~nm, the IFCs collapse into nearly parallel lines, characteristic of the canalization regime, whereas at $\lambda_0$ = 5000~nm and 6000~nm the contours become closed, indicating the elliptic regime. Dashed blue lines are analytically calculated IFCs in the small thickness limit, showing excellent agreement of the analytical model and experimental FFTs. The observation of the polariton topological transition is further well confirmed by the simulated $\lvert E_z \rvert$ maps shown in Fig.~\hyperref[fig:fig2]{2e}, as well as by the corresponding momentum-space ($k$-space) images presented in Fig.~\hyperref[fig:fig2]{2f} together with analytical calculations (blue dashed lines).

\subsection{Canalization wavelength engineering \textit{via} flake thickness}

\begin{figure}[h!]
\captionsetup{skip=12pt}
    \centering
    \includegraphics[width=1.0\linewidth]{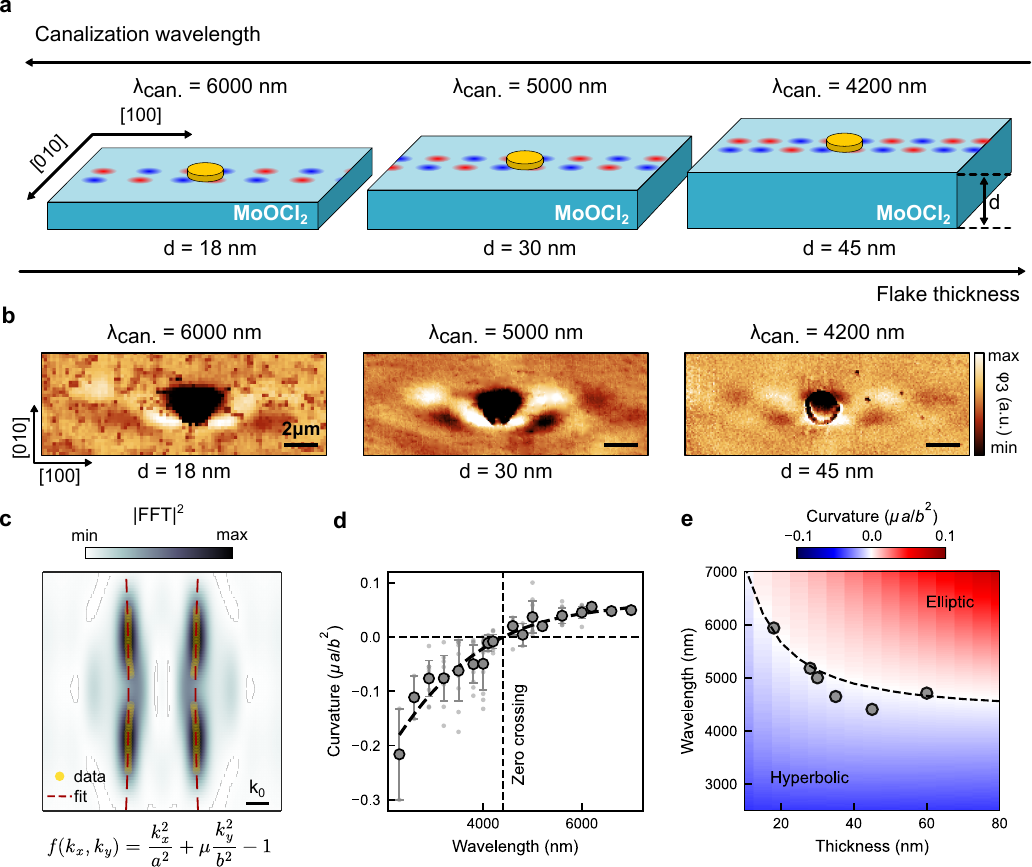}
    \caption{\textbf{Tailoring canalization wavelength \textit{via} flake thickness}. \textbf{a} Schematic illustrating the shift of the canalization wavelength toward lower values with increasing flake thickness. \textbf{b} Real-space sSNOM phase maps (third demodulation order) showing canalized wavefronts launched by the gold disk nanoantenna for flakes of different thicknesses. \textbf{c} Fourier transform of the canalised wavefront, fitted with a conic function (red dashed lines) to extract the curvature values. \textbf{d} The curvature plot illustrates the transition of the supported modes from hyperbolic to canalised and subsequently to elliptical. Small grey dots represent curvature values extracted from the experimental data, while the larger grey dots correspond to their averaged values. The large grey dots are fitted with a polynomial curve to identify the zero-crossing position (vertical dashed line), which marks the expected canalization wavelength at approximately 4400 nm. \textbf{e} Plasmon-polariton dispersion regimes with canalization window (white area) obtained from experiment (grey dots) and from simulations (dashed line). The blue and red regions correspond to hyperbolic and elliptic regimes, respectively.}
    \label{fig:fig3}
\end{figure}

As anticipated in Fig.~\hyperref[fig:fig1]{3e}, the canalization wavelength can be controlled by tuning the flake thickness. Fig.~\hyperref[fig:fig3]{3a} schematically illustrates this effect. As the flake thickness increases, the wavelength at which canalization occurs shifts toward shorter values. To experimentally confirm this theoretical prediction, we carry out near-field experiments at frequencies both above and below the topological transition for several flakes of different thickness. In these experiments, gold disks are used as launchers instead of nanorods. The disks excite the same modes as the nanorods, as the launcher geometry for subwavelength scatterers does not significantly alter the observed SPP pattern. Additional experimental data obtained using both gold nanorod and disk launchers are provided in Supplementary Information \ref{SNOM_maps}.

The experimental observations confirming the blueshift of the canalization wavelength with increasing thickness are shown in Fig.~\hyperref[fig:fig3]{3b}. For a 18~nm-thick $\mathrm{MoOCl}_{2}$ flake, the canalization wavelength is observed at 6000~nm. Increasing the thickness to 30~nm shifts the canalization wavelength to 5000~nm, while for a thickness of 45~nm, canalization is observed at 4200~nm. In order to quantitatively evaluate the shift of the canalization frequency we fit the experimental k-space images with the conic function $f(k_x,k_y) = k_x^2/a^2 + \mu k_y^2/b^2 - 1$, which can represent both elliptical ($\mu = 1$) and hyperbolic ($\mu = -1$) IFCs (see an example in Fig. ~\hyperref[fig:fig3]{3c}). In order to perform the fit we have to reduce each image to a set of discrete k-space points: we perform this task by binarization of the FFT images, followed by detection of connected regions. From each of these regions, we extract a continuous line from the row-by-row weighted average of the data (for more details see Supplementary Information \ref{Exp_treat}). From the conic fit, we compute the vertex curvature at the $k_x=0$ point as $\kappa = \mu a/b^2$, which can be used as a clear signature of the topological transition, as $\kappa = 0$ represents straight line IFCs. From measurements taken on a wide range of wavelengths we can produce a curvature vs. wavelength curve (Fig. ~\hyperref[fig:fig3]{3d}), which we fit with a third-order polynomial to extract the zero crossing point, representing our retrieved canalization wavelength (see Supplementary Information \ref{Exp_canal_freq}). We carry out a similar procedure by using the same conic function to fit the analytically calculated IFCs for films with different thicknesses (see Supplementary Information \ref{Exp_canal_freq} for more details). As a result, we obtain the image in Fig. ~\hyperref[fig:fig3]{3e}, from which we can trace the evolution of the zero-crossing as highlighted by the black dashed line. The fact that the canalization frequency moves with the film thickness is not unexpected, as the shape of the IFCs for a hyperbolic film depend not only on the dielectric function, but also on the layer thickness. By looking at the equation for the theoretical IFCs (Supplementary Information \ref{Theo_canal_freq}), additional insight into the thickness dependence of the canalization condition can be obtained. Assuming a frequency-independent dielectric response, with one in-plane permittivity component remaining deeply negative and the other negative but closer to zero, the canalization wavelength is expected to increase linearly with thickness. This follows from the analytical expression, where the slab thickness and free-space wavevector appear only through their product, so that the canalization condition is preserved by scaling thickness and wavelength by the same factor (see Supplementary Information \ref{Theo_canal_freq}). In MoOCl$_2$, however, the opposite trend is observed experimentally and theoretically: increasing the flake thickness leads to a reduction of the canalization wavelength. This behavior is mainly driven by the dispersion of the $[010]$ permittivity, which governs the evolution of the IFC shape across the topological transition. Since the permittivity enters the canalization condition multiplied by the flake thickness $d$, an increase in the absolute value of $\varepsilon_y$ (i.e., a more negative response) requires a smaller thickness to satisfy the condition. As a result, the thickness dependence of the canalization wavelength qualitatively follows the Drude-like dispersion of the $[010]$ axis, explaining the trend reported in Fig.~\hyperref[fig:fig3]{3e}. In contrast, the permittivity along the $[100]$ direction remains strongly negative throughout the relevant spectral range and therefore behaves almost as a perfect electric conductor, contributing only weakly to the thickness-induced shift of the canalization condition.

From sSNOM experiments on flakes between \SI{18}{\nano \meter} and \SI{60}{\nano \meter}, we obtain the experimental data plotted as dots in Fig. ~\hyperref[fig:fig3]{3e}, confirming that the canalization wavelength can be tuned over more than $\SI{1}{\micro \meter}$, corresponding to a variation $>25\%$ of the central wavelength by adjusting the flake thickness. We also show that in a model Lorentz-Lorentz phonon-polariton material, canalization naturally emerges only within a narrow spectral. As a consequence, the achievable thickness-dependent tuning remains intrinsically limited by the restricted bandwidth over which canalization exists (see Supplementary Information \ref{Theo_canal_freq}). In contrast, the Drude-like response of MoOCl$_2$ enables canalization across a much broader spectral range, allowing substantially larger shifts of the canalization wavelength with thickness. Importantly, such a high tuning bandwidth goes well beyond what has been achieved in phonon-polariton systems, where tuning via twist angle in bilayer \cite{hu2020topological, duan2020twisted, chen2020configurable} and trilayer \cite{duan2023multiple} heterostructures or by substrate engineering \cite{duan2025canalization, zhu2025multiple} was constrained to $<9\%$ of the central wavelength. Larger tunability has been reported for PhPs at THz frequencies \cite{obst2023terahertz}, yet it is still limited to less than $14\%$ of the central energy.

\section{Conclusions}

In summary, we have demonstrated that strongly directional, diffractionless plasmon-polariton propagation can arise naturally in a low-symmetry van der Waals crystal by exploiting its intrinsic elliptical-to-hyperbolic topological transition, rather than relying on external patterning, twist engineering, or artificial metastructures. This establishes a conceptually simple route to diffraction-free plasmon transport that is robust against fabrication imperfections and compatible with atomically thin, planar architectures. 

An important implication of this work is that canalization in plasmonic systems does not need to be confined to a narrowly defined operating point. In $\mathrm{MoOCl}_{2}$, the gradual Drude dispersion enables a finite spectral window in which highly directional propagation persists, while thickness provides an efficient knob to position this window across the mid-infrared. This combination of broadband response and tunability suggests new strategies for matching polariton propagation to external constraints, such as molecular vibrational bands, detector sensitivities, or on-chip optical components.

Beyond $\mathrm{MoOCl}_{2}$, our findings motivate a broader search for natural canalization in other anisotropic conductors and semimetals, particularly those exhibiting orbital-selective or symmetry-driven electronic anisotropy. Materials hosting multiple zero crossings or coexisting in-plane transitions could support cascaded or multi-channel canalized propagation. Furthermore, integrating such crystals with electrostatic gating, phase-change substrates, or strain fields may enable real-time steering and reconfiguration of polariton beams without lithographic modification.

\subsection*{Acknowledgments}

A.M. acknowledges funding from the European Union’s Horizon Europe research and innovation programme under Marie Skłodowska-Curie grant agreement no. 101146874. A.A. acknowledges funding from the European Union (ERC-2025-POC 2Dchiral N.101248056). We acknowledge support from the ERC (Complexplas, 3DPrintedoptics) (F.A., B.F. and H.G.), DFG (SPP1391 Ultrafast Nanooptics, CRC 1242 “Non-Equilibrium Dynamics of Condensed Matter in the Time Domain” project no. 278162697-SFB 1242, Germany’s Excellence Strategy EXC 2089/1–390776260, Emmy Noether Program TI 1063/1) (F.A., B.F. and H.G.), BMBF (Printoptics) (F.A., B.F. and H.G.), BW Stiftung (Spitzenforschung, Opterial) (F.A., B.F. and H.G.), Carl-Zeiss Stiftung (F.A., B.F. and H.G.) and DFG (GRK2642 Photonic Quantum Engineers). F.A. acknowledges Florian Mangold (4th Physics Institute and Research Center SCoPE, University of Stuttgart) for valuable discussions and for providing the phase evolution video of canalization (see Supplementary Videos).

\section*{Competing Interests Statement}

The authors declare no competing interests.

%TC:ignore
\section*{Methods}\label{sec11}

\subsection*{Sample fabrication}

MoOCl$_2$ flakes were obtained by mechanical exfoliation from the corresponding bulk crystal (Nanjing MKNANO Tech. Co., Ltd.). Flakes investigated in the far-field experiments were deposited on SiO$_2$, while for the near-field experiments we used Si/SiO$_2$ (285 nm) substrates (Wafer University). Flakes with suitable ranges of thicknesses were selected for the launcher fabrication via the flake colour. The thicknesses were then confirmed by AFM measurements. The nanostructures were fabricated by a standard electron-beam lithography followed by metal deposition and lift-off. A layer of poly(methyl methacrylate) resist, PMMA 950K A4, was spin-coated onto the substrate at 4000 rpm for 60 s and subsequently baked at 180 °C for 3 min. The resist was patterned using 50 kV electron-beam lithography with a 50 µm aperture. After exposure, the sample was developed for 60 s in a 1:3 mixture of methyl isobutyl ketone and isopropyl alcohol. The development was stopped by rinsing the sample in isopropyl alcohol, followed by drying under a nitrogen gas flow. A 2 nm chromium adhesion layer and a 45 nm gold layer were then deposited by electron-beam evaporation. Finally, lift-off was performed by immersing the sample in acetone for more than 3 h, leaving the patterned gold nanostructures on the flakes.

\subsection*{Fourier Transform Infrared spectroscopy}

Polarization-resolved reflectivity and transmission infrared spectra were acquired on the sample using a Bruker Vertex 80 microscope Hyperion 3000 equipped with a 36X Cassegrain objective. For each crystallographic axis of MoOCl$_2$, the reflectance and transmittance were measured under near-normal incidence for a number of flakes with different thicknesses. The in-plane dielectric function was extracted by fitting the measured spectra with a 3-layer model (vacuum-MoOCl$_2$-SiO$_2$) analogously to what was performed in Ref. \cite{Melchioni2025}. For simplicity, we ignored the off-angle illumination introduced by the Cassegrain objective. Transmission measurements were fitted only up to \SI{5}{\micro \meter} because of the onset of absorption of the glass substrate. The final dielectric function was obtained by fitting the average dielectric function with two Drude-Lorentz model (one for each crystal axis). Before fitting, the thickness of each flake was fixed by fitting the optical response (reflection and transmission) in the visible along both axes with the dielectric function reported in Ref. \cite{Melchioni2025} (see also Supporting Information \ref{Eps_extraction}). The fitting was performed with the layer thickness as the only free parameter. Infrared spectra were acquired using a broadband light source in conjunction with a CaF$_2$ near-infrared beamsplitter, covering the spectral range from 0.7 to 8.3~\textmu m. Visible spectra were acquired with a custom-made microscope using a quartz tungsten-halogen Lamp (Thorlabs, QTH10) as the light source. Light was focused and collected by a pair of 20x long working distance objectives (Mitutoyo) and directed to a fiber coupled spectrometer (Thorlabs, CCT10). Before collection, a 4f system and an iris were used to cut the re-imaged image plane to the area of the flake we intended to measure. A polarizer (Thorlabs, WP25M-UB1) in conjunction with a broadband half waveplate (Thorlabs, AQWP05M-580) were used to measure along the different axes of the crystal. By rotating the half waveplate and collecting angle-dependent reflectivity spectra, we determine the [100] and [010] axes as the ones with, respectively, the maximum and minimum integrated reflectance in the range 700 - \SI{850}{\nano \meter}.

\subsection*{sSNOM}
Near-field maps presented in Fig.~\hyperref[fig:fig2]{2c} were acquired using a commercial room-temperature sSNOM from Neaspec GmbH coupled to a broadband tunable laser source from Stuttgart Instruments GmbH covering 1450–20000 nm wavelength range. In the present experiments, the laser wavelength was tuned from 2300 nm to 7000 nm to selectively excite specific plasmon-polaritonic modes. The laser beam was focused by an off-axis parabolic mirror (incident angle $\approx \SI{60}{\degree}$) onto the apex of a metal-coated AFM tip (NCLPt, Pt/Ir coating). Light is p-polarized in order to maximize the projection along the tip’s vertical axis. The tip was operated in tapping mode, with the cantilever oscillating at 193 kHz, and the tapping amplitude was adjusted depending on the wavelength to ensure operation strictly in the near-field regime and to maximize signal-to-noise ratio.\\
The sample, consisting of gold launchers (disks and nanorods) fabricated on a $\mathrm{MoOCl}_{2}$ flake, was raster-scanned beneath the AFM tip. The tip acts as a localized near-field source, providing the momentum required to launch plasmon-polaritons (PPs) on the sample surface. These modes propagate across the $\mathrm{MoOCl}_{2}$ flake and are scattered by the gold launcher, or conversely, the launcher scatters modes launched by the tip. The resulting interference between light scattered from the tip and the launcher generates a modulated far-field signal at the detector. The amplitude and phase of this modulation depend sensitively on the tip–launcher distance, enabling high-resolution imaging of the spatial distribution and propagation characteristics of the plasmon-polaritons. To isolate the near-field contribution from background far-field scattering, the detector signal was demodulated at higher harmonics (n$\mathrm{\Omega}$, n = 3) of the cantilever oscillation frequency.\\
Prior to focusing on the sample, half of the laser beam was redirected to a pseudo-heterodyne interferometer via a beam splitter. Light scattered by the tip was then recombined with the interferometer reference, allowing retrieval of both amplitude- and phase-resolved maps of the electromagnetic field at the sample surface by properly setting the mirror oscillation amplitude \cite{ocelic2006pseudoheterodyne}. To maximize plasmon-polariton scattering efficiency, the $\mathrm{MoOCl}_{2}$ flake was rotated such that its dielectric axis was aligned with the laser incidence direction (Fig.~\hyperref[fig:fig2]{2b}).

\subsection*{Numerical simulations}

Electromagnetic simulations were performed with a commercial solver (CST Studio) in the frequency domain. The dielectric function of MoOCl$_2$ was extracted from the experimental optical spectra as detailed above. For the dipole simulations, we placed a near-field source above (\SI{50}{\nm}) a MoOCl$_2$ slab.
The near-field source was obtained by inserting a discrete port in a vacuum gap inside a perfect electric conductor (PEC) cylinder. As CST does not support open boundaries for anisotropic media, a vacuum gap was inserted between the end of the MoOCl$_2$ slab and the open boundary conditions. Simulations of the PPs launched by disks and rods (modeled as PEC solids) were carried out with a plane-wave sourced with a \SI{60}{\degree} tilt from the normal direction.

Calculations of the imaginary part of the reflection coefficients were performed using a 4×4 transfer matrix formalism \cite{passler2017generalized} for the homogenous vacuum-MoOCl$_2$-SiO$_2$ stack. The two-dimensional maps were obtained by rotating the MoOCl$_2$ dielectric function around the z-axis and computing the reflection coefficients at each angle.

\section*{Data availability}
The data supporting the findings of this study are available from the corresponding author upon reasonable request.

\section*{Ethics declarations}
The authors declare no competing interests.

\section*{Author Contributions}

F.A. and A.M. performed the measurements. L.N. fabricated the samples. A.M. and G.V. carried out numerical simulations. F.A. and A.M. analyzed the data. B.F., H.G. and A.A. supervised the project. F.A. and A.M. wrote the manuscript with input from all authors. All authors contributed to the discussion of the results.

\clearpage
\section*{Supplementary Information}
\addcontentsline{toc}{section}{Supplementary Information}

\setcounter{figure}{0}
\setcounter{table}{0}
\setcounter{equation}{0}
\setcounter{section}{0}

\renewcommand{\thefigure}{S\arabic{figure}}
\renewcommand{\thetable}{S\arabic{table}}
\renewcommand{\theequation}{S\arabic{equation}}
\renewcommand{\thesection}{S\arabic{section}}

\section{Extraction of mid-IR in-plane dielectric function}\label{Eps_extraction}

\begin{figure}[h!]
    \centering
    \includegraphics[width=1.0\linewidth]{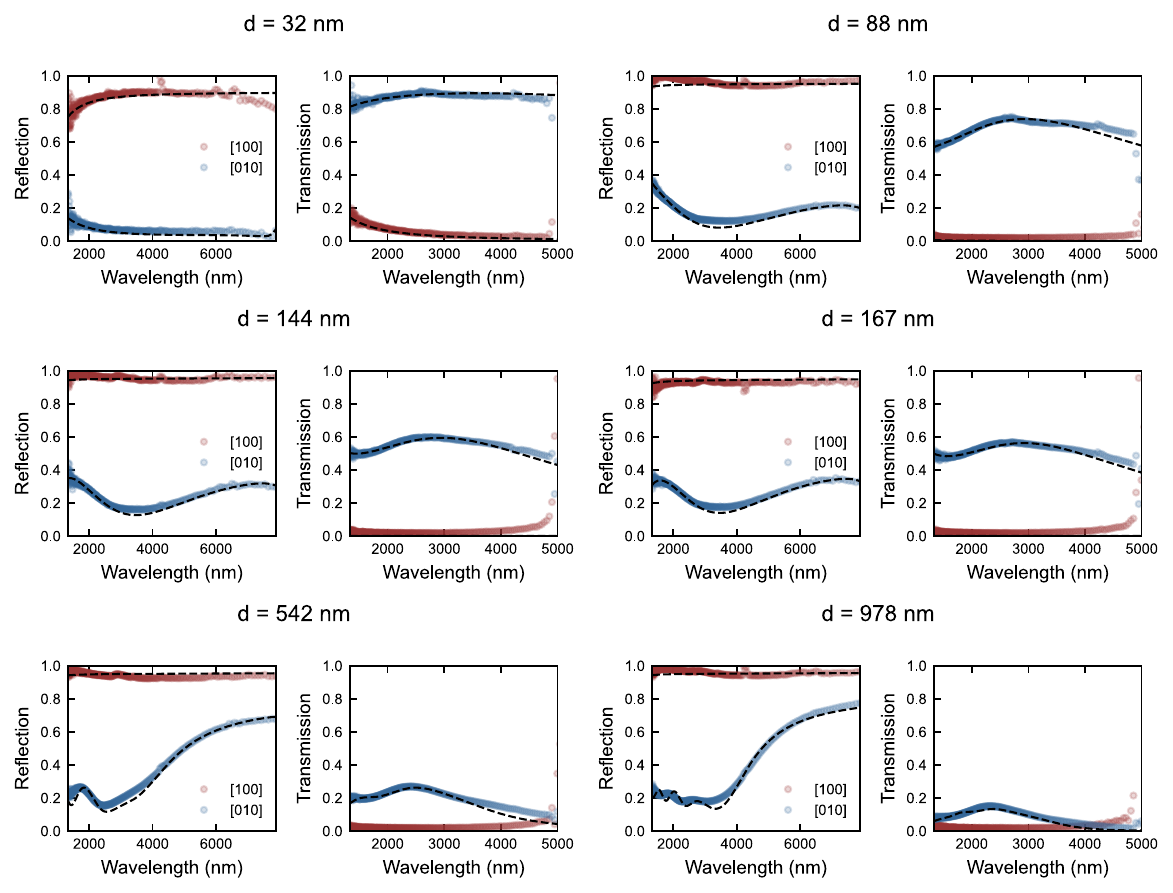}
    \caption{\textbf{Extraction of the in-plane dielectric function.} Fits of FTIR reflection and transmission spectra (dashed black lines) with a three layer model from which the in-plane dielectric function is extracted.}
    \label{fig:Fit_thickness}
\end{figure}

The in-plane dielectric function was extracted similarly to Ref. \cite{Melchioni2025} by fitting experimental reflection and transmission spectra with a 3-layer model (vacuum-MoOCl$_2$-SiO$_2$). Polarization-resolved reflectivity and transmission infrared spectra acquired on several MoOCl$_2$ flakes taken with a FTIR microscope are shown in Fig. \ref{fig:Fit_thickness}. The two axes are readily identified by the minimum and maximum reflection points while rotating the polarizer (see also Methods section). For simplicity, we neglect here the tilted illumination provided by the Cassegrain objective and assume normal incidence. The three-layer system is modeled by the following equations for the reflection and transmission coefficients \cite{dressel2002electrodynamics}:

\begin{gather} \label{eq1}
r = \frac{r_{12} + r_{23}e^{2i\delta}}{1+r_{12}r_{23}e^{2i\delta}}\\
t = \frac{t_{12} t_{23}e^{i\delta}}{1+r_{12}r_{23}e^{2i\delta}}
\end{gather}

where the $r_{ij}$ and $t_{ij}$ are Frensel coefficients of single interfaces:

\begin{gather*} \label{eq2}
\delta = 2\pi dn/\lambda\\
r_{ij} = \frac{n_i-n_j}{n_i+n_j}\\
t_{ij} = \frac{2n_j}{n_i+n_j}
\end{gather*}

\begin{figure}[t!]
    \centering
    \includegraphics[width=0.75\linewidth]{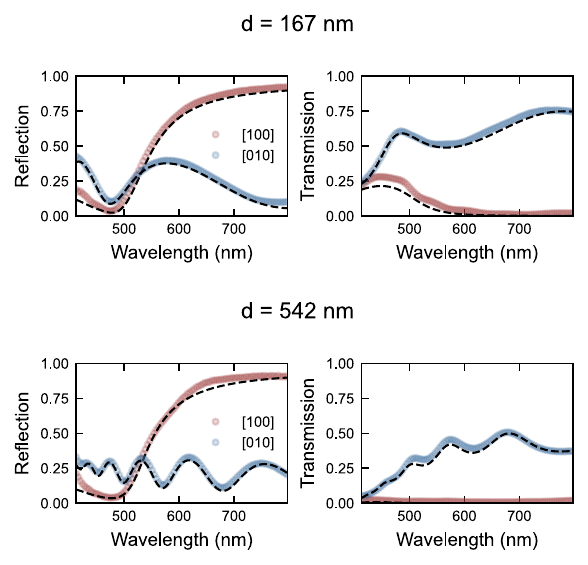}
    \caption{\textbf{Determination of flake thickness via visible spectroscopy.} Examples of the determination of flake thicknesses by fitting optical spectra in a range where the dielectric function is known \cite{Melchioni2025}.}
    \label{fig:Fit_thickness_vis}
\end{figure}

Here $d$ and $n$ are the MoOCl$_2$ thickness and complex refractive index, respectively, and $\lambda$ is the wavelength. The indexes correspond to $1$ for the material above the flake (air, $n=1$), $2$ for the MoOCl$_2$ layer, and $3$ for the SiO$_2$ substrate. The observed reflection and transmission are then calculated as $R=|r|^2$ and $T=|t|^2/n_{3}$. Each flake thickness is determined from optical spectra taken in the visible as detailed in Ref. \cite{Melchioni2025}, where the dielectric function is fixed and the thickness is left as the only fitting parameter (Fig. \ref{fig:Fit_thickness_vis}).

\begin{figure}[h!]
    \centering
    \includegraphics[width=1.0\linewidth]{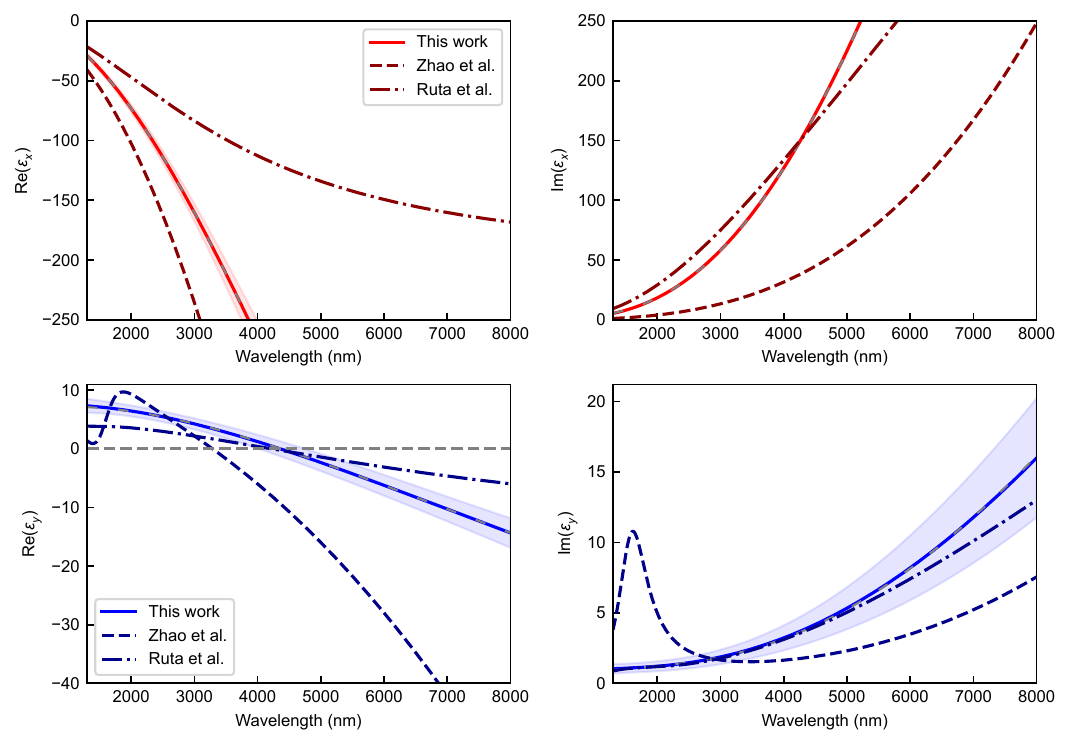}
    \caption{\textbf{Mid-IR dielectric function comparison.} Comparison of the extracted mid-IR complex dielectric function along both crystal directions with previously proposed models \cite{zhao2020highly, ruta2024good}.}
    \label{fig:Fit_dielectric_function_comparison}
\end{figure}

For the extraction of the mid-IR parameters, we fit the dielectric function of both the MoOCl$_2$ $[100]$ and $[010]$ axes with a Lorentz-Drude model:

\begin{equation}\label{DrudeLor}
    \varepsilon(\omega) = \varepsilon_{\infty} - \frac{\omega_p^2}{\omega(\omega + i\gamma)} + \frac{\omega_{p,j}^2}{\omega_{0,j}^2-\omega^2 - i\gamma_j\omega}
\end{equation}

 The parameters in Eq.\eqref{DrudeLor} are the ones used to minimize the fit to the experimental data. For a robust estimation of the dielectric function parameters, we fit reflection and transmission spectra along a single axis at the same time with the same Drude-Lorentz values. We do this by stitching the reflection and transmission experimental data, which we fit with the concatenation of the calculated spectra from the three layer system. The extracted Drude Lorentz parameters are reported in table \ref{Si_Tab_Fit_Parameters}, while a comparison of the dielectric function with the one calculated by DFT \cite{zhao2020highly} and extracted via reflection experiment of a single thick flake \cite{ruta2024good} is shown in  Fig. \ref{fig:Fit_dielectric_function_comparison}.

 \begin{table}[!h]
 \captionsetup{skip=10pt}
    \centering
    \begin{tabular}{c||c|c|c|c|c|c|c|c|c}
        Axis & $\varepsilon_\infty$ & $\omega_p$ & $\gamma$ & $\omega_{0,1}$ & $\omega_{p,1}$ & $\gamma_1$   \\
        \hline
        $\unit{[100]}$ &  0 & 5.63 & 0.15 & 0 & 0 & 0 \\
        $\unit{[010]}$ & 7.85 & 0.91 & 0.11 & 1.15 & 1.30 & 1.99\\
    \end{tabular}
    \caption{\textbf{Values of the parameters for the mid-IR dielectric function extracted from the fit}. All the values are reported in eV, except for $\varepsilon_\infty$ which is a dimensionless parameter.}
    \label{Si_Tab_Fit_Parameters}
\end{table}

\begin{figure}[h!]
    \centering
    \includegraphics[width=0.75\linewidth]{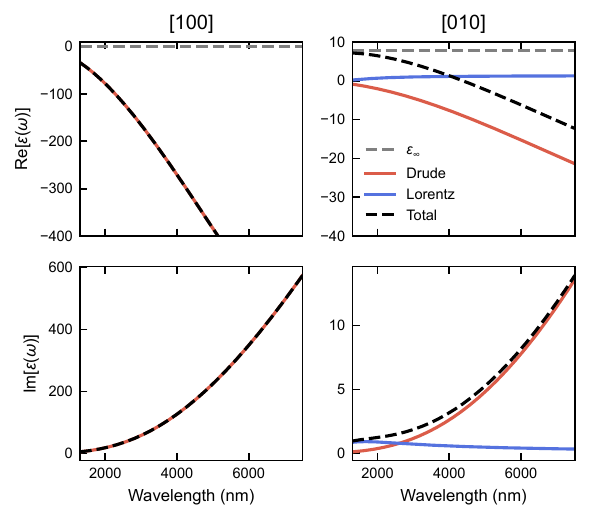}
    \caption{\textbf{Mid-IR dielectric function decomposition} Contribution from the Drude and Lorentz terms to the extracted dielectric function of MoOCl$_2$ in the IR range.}
    \label{fig:dielectric_function_components}
\end{figure}

The decomposition of the extracted dielectric function into its constituent contributions (Fig. \ref{fig:dielectric_function_components}) highlights the distinct physical origins of the optical response along the two crystallographic directions. In the [100]-direction, the permittivity is well described by a single Drude term, indicating a predominantly free-carrier response with no significant interband contributions in the spectral range of interest. In contrast, the [010]-direction requires the inclusion of an additional Lorentz oscillator to accurately capture the low transmission at lower wavelengths in the investigated range.

\begin{figure}[ht!]
    \centering
    \includegraphics[width=1\linewidth]{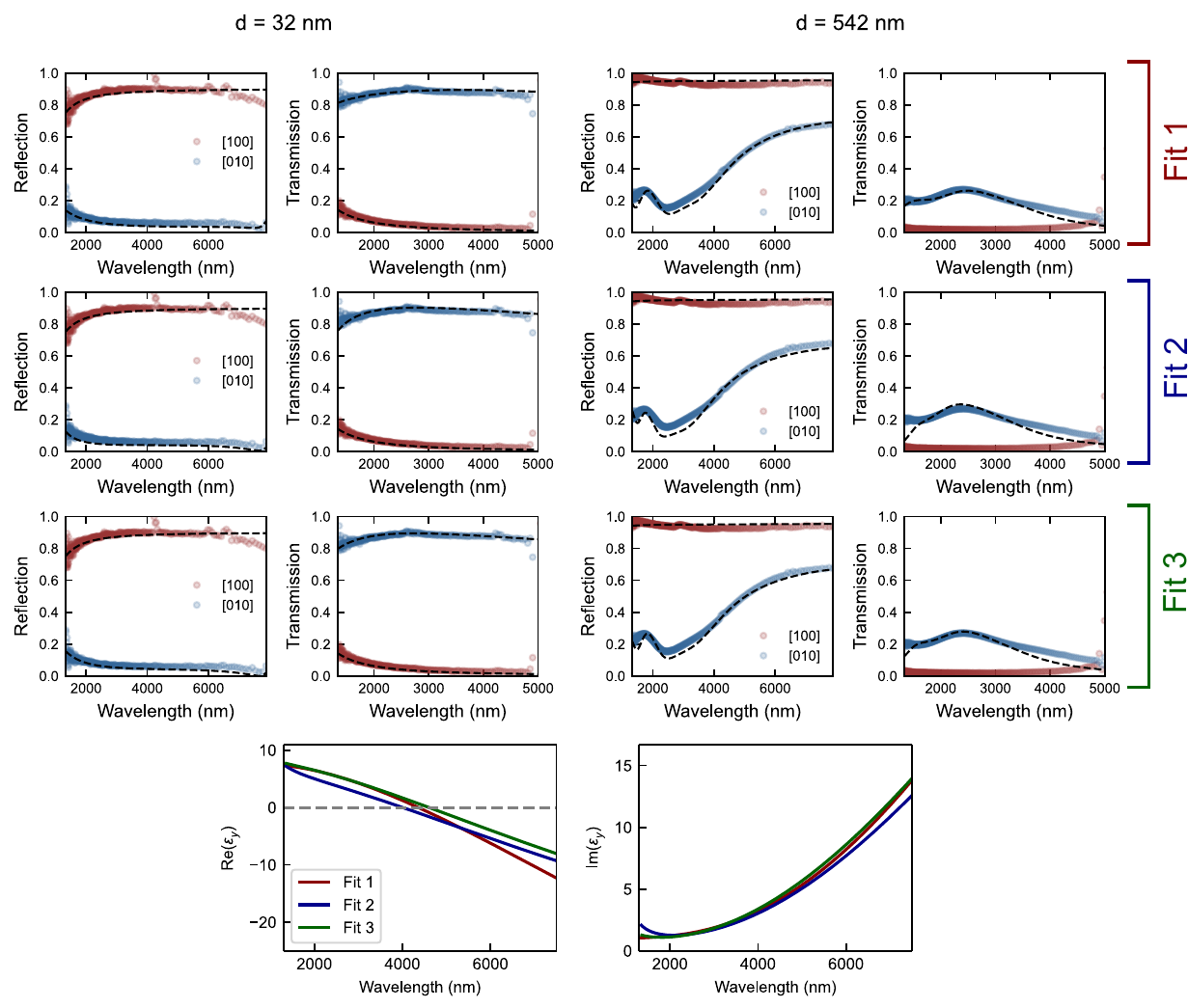}
    \caption{\textbf{Sensitivity of the [010] dielectric function fitting.} Reflection and transmission spectra along the [100] and [010] directions for two representative flake thicknesses (32 and 542 nm) are shown together with three fits obtained from different starting parameters. Bottom panels show the corresponding extracted real and imaginary parts (obtained from the average of all the flakes thicknesses as in Fig. \ref{fig:Fit_thickness}) of $\varepsilon$ along the [010] axis, illustrating the variability arising from fitting degeneracy.}
    \label{fig:dielectric_function_sensitivity}
\end{figure}

We also performed a sensitivity analysis of the dielectric function fitting. For the [100] direction, $\mathrm{Re}(\varepsilon_x)$ is strongly negative in the spectral range of interest, so the response is effectively metallic and the fit is well constrained by a single Drude term, leading to negligible impact on the extracted IFCs. In contrast, the [010] direction allows for greater flexibility, as different parameter combinations can reproduce the reflection and transmission data with comparable accuracy (Fig. \ref{fig:dielectric_function_sensitivity}, Fit 1 corresponds to the one used in the manuscript with parameters reported above). In particular, variations in $\varepsilon_\infty$ can be partially compensated by a redistribution between the Drude and Lorentz contributions. To assess this, we performed additional fits starting from different $\varepsilon_\infty$ values and extracted the corresponding dielectric functions. The resulting variation is illustrated in the new figure, where representative fits and the corresponding extracted permittivities are shown for two flake thicknesses, with the associated fit parameters (for Fit \#2 and \#3) reported in Table \ref{Si_Tab_Fit_Parameters_sensitivity}.

 \begin{table}[!h]
 \captionsetup{skip=10pt}
    \centering
    \begin{tabular}{c||c|c|c|c|c|c|c|c|c}
        Fit \# ([010]) & $\varepsilon_\infty$ & $\omega_p$ & $\gamma$ & $\omega_{0,1}$ & $\omega_{p,1}$ & $\gamma_1$   \\
        \hline
        $1$ &  4.32 & 0.84 & 0.14 & 1.13 & 1.49 & 0.35 \\
        $2$ & 6.99 & 0.88 & 0.14 & 1.29 & 1.49 & 0.93\\
    \end{tabular}
    \caption{\textbf{Values of the parameters for the [010] mid-IR dielectric function extracted from the additional fits \#2 and \#3 in Fig. \ref{fig:dielectric_function_sensitivity}}. All the values are reported in eV, except for $\varepsilon_\infty$ which is a dimensionless parameter.}
    \label{Si_Tab_Fit_Parameters_sensitivity}
\end{table}

\begin{figure}[ht!]
    \centering
    \includegraphics[width=1\linewidth]{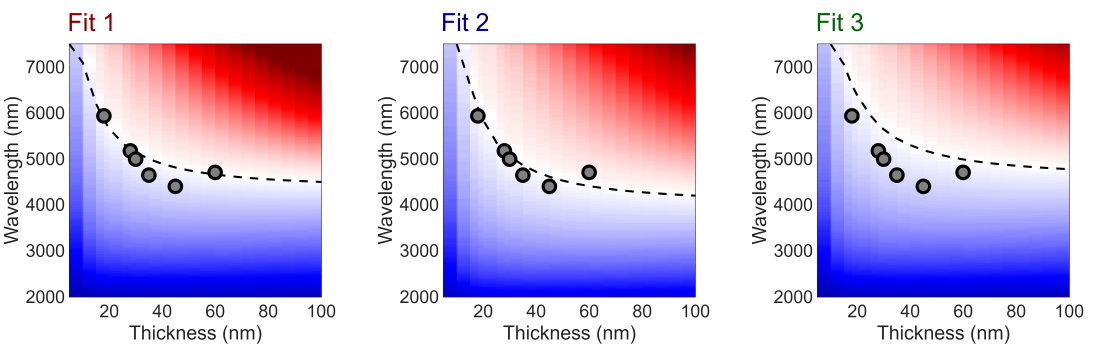}
    \caption{\textbf{Comparison of expected shift of the canalization wavelength with flake thickness for different fitted [010] permittivity.} IFCs curvature evolution as a function of wavelength and thickness for the different fits shown in Fig. \ref{fig:Fit_dielectric_function_comparison}.}
    \label{fig:curvature_sensitivity}
\end{figure}

The sensitivity of the polariton IFCs to the different dielectric-function fits is shown in Fig. \ref{fig:curvature_sensitivity}. For each fitted permittivity, we recalculated the IFCs, extracted the curvature using the same conic fitting procedure employed in the main text, and determined the curvature zero-crossing wavelength as a function of thickness. The calculated curvature maps and zero-crossing lines show that the qualitative behavior is robust across the different parameterizations: all fits reproduce a thickness-dependent redshift of the canalization wavelength with a similar trend of the hyperbolic-to-elliptical evolution of the IFCs. However, the absolute spectral position of the canalization curve depends on the fitted $\varepsilon_y(\omega)$, with Fit~3 producing a more pronounced redshift. Thus, the uncertainty in the dielectric-function fitting affects the precise calculated canalization wavelength, but not the main conclusion that intrinsic canalization persists and remains thickness-tunable over the experimentally observed range.

\newpage

\section{Role of the polarization for the efficient excitation of plasmon polaritons}\label{Polaritons_polarization}

\subsection{Simulations of launched polaritons under different polarization and rod orientation}

\begin{figure}[h!]
    \centering
    \includegraphics[width=0.75\linewidth]{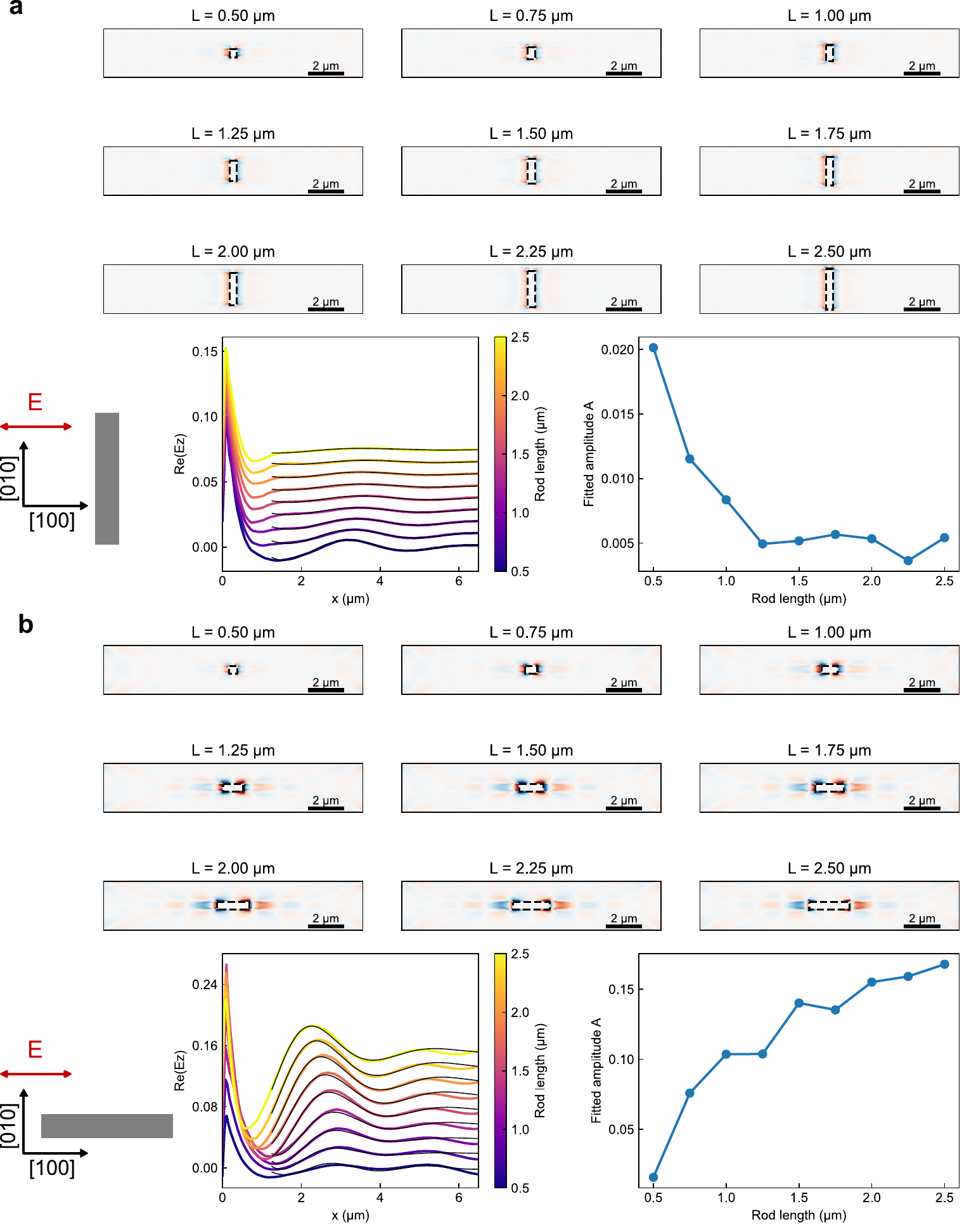}
    \caption{ \textbf{Simulated polariton launching in a 45 nm flake at the canalization frequency with polarization along the [100] direction.} Simulated $E_z$ maps launched by a rod of various lengths and fixed width. The rods are aligned with the [010] direction in \textbf{a} and with the [100] direction in \textbf{b}. Electric field profiles are fitted with an oscillating decay curve and the extracted amplitude is reported as a function of the rod length.}
    \label{fig:polarization_dependance_sim}
\end{figure}

\begin{figure}[h!]
    \centering
    \includegraphics[width=0.75\linewidth]{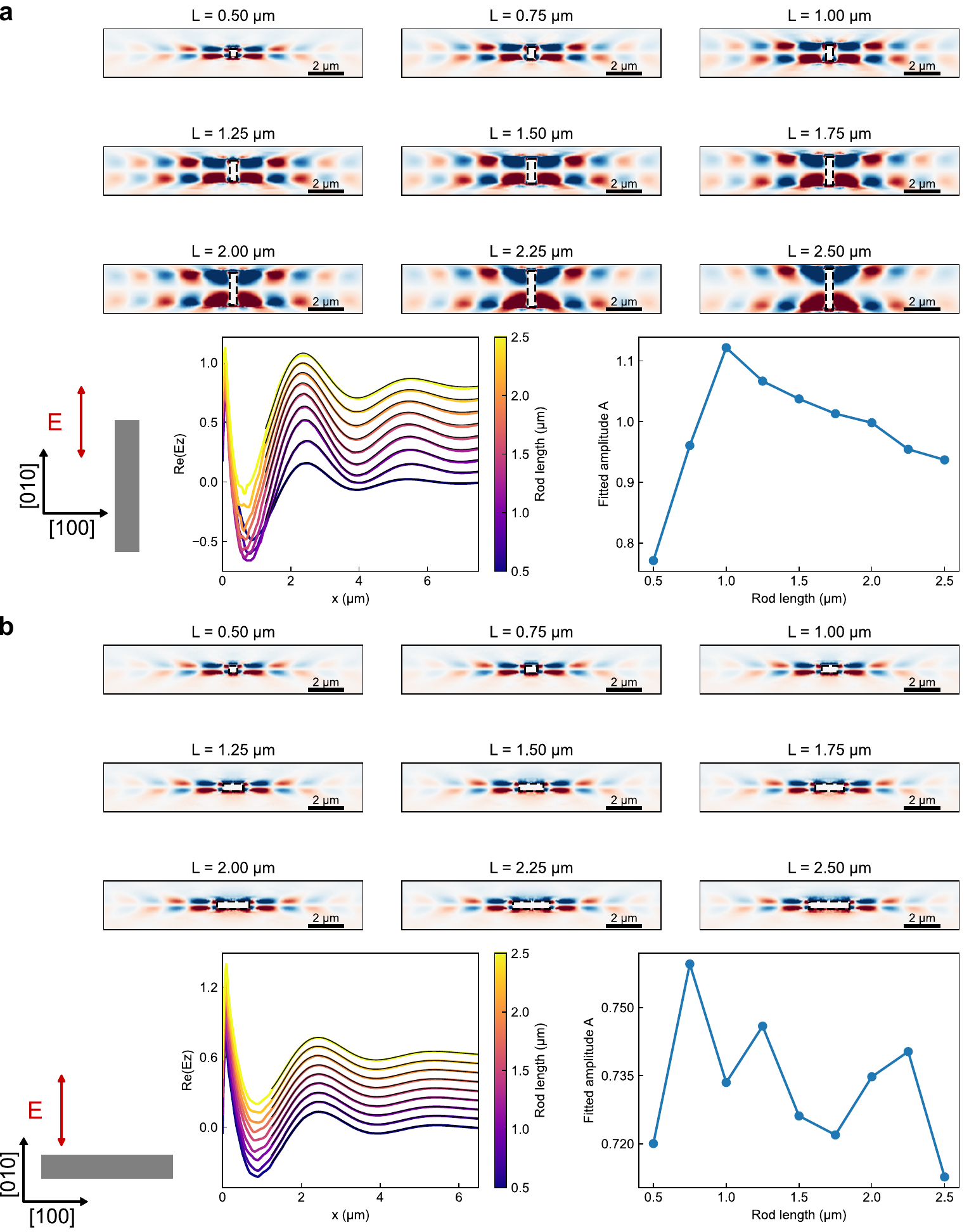}
    \caption{ \textbf{Simulated polariton launching in a 45 nm flake at the canalization frequency with polarization along the [010] direction.} Simulated $E_z$ maps launched by a rod of various lengths and fixed width. The rods are aligned with the [010] direction in \textbf{a} and with the [100] direction in \textbf{b}. Electric field profiles are fitted with an oscillating decay curve and the extracted amplitude is reported as a function of the rod length.}
    \label{fig:polarization_dependance_sim2}
\end{figure}

To clarify the role of the excitation geometry and the associated launching efficiency, we performed additional simulations at the canalization frequency for a 45 nm thick layer (Fig. \ref{fig:polarization_dependance_sim} and Fig. \ref{fig:polarization_dependance_sim}2). For simplicity, we considered normal-incidence excitation with a linearly polarized plane wave and analyzed four configurations combining polarization along the [100] and [010] directions with rod antennas (modeled as perfect electric conductors with lengths between 450 nm and 2500 nm and fixed width of 450 nm) oriented along either [100] or [010]. For each configuration, we shows the simulated $E_z$ field at the MoOCl$_2$ surface for each rod length and a sketch of the rod and polarization the exciting plane wave. For each rod length, the polariton profile was extracted from the simulated electric field (curves are vertically offset for clarity) and fitted with a decaying oscillatory exponential ($f(x)=Ae^{-\alpha x}\cos{kx+\phi}$) to quantify the launching efficiency via the extracted amplitude. We find that when the rod is aligned along [010] and the incident polarization is along [100], the excitation remains weak for all rod lengths (Fig. \ref{fig:polarization_dependance_sim}a). When both rod and polarization are aligned along [100], the efficiency increases monotonically with rod length, indicating the onset of a resonance at larger sizes (Fig. \ref{fig:polarization_dependance_sim}b). The highest launching efficiency is obtained when both rod and polarization are along [010], where a clear maximum is observed around a rod length of $\sim$1.1~$\mu$m, consistent with a resonance-like behavior (Fig. \ref{fig:polarization_dependance_sim2}a). Importantly, the propagation profile remains unchanged aside from intensity variations, indicating that antenna effects do not distort the canalized mode in the simulated conditions. In the experimental configuration (rod along [100], polarization along [010]), we observe robust and efficient excitation that is only slightly weaker than the resonant case and largely independent of rod length (Fig. \ref{fig:polarization_dependance_sim2}b).

\subsection{Understanding the advantage of using polarization aligned with the [010] direction}

\begin{figure}[h!]
    \centering
    \includegraphics[width=0.75\linewidth]{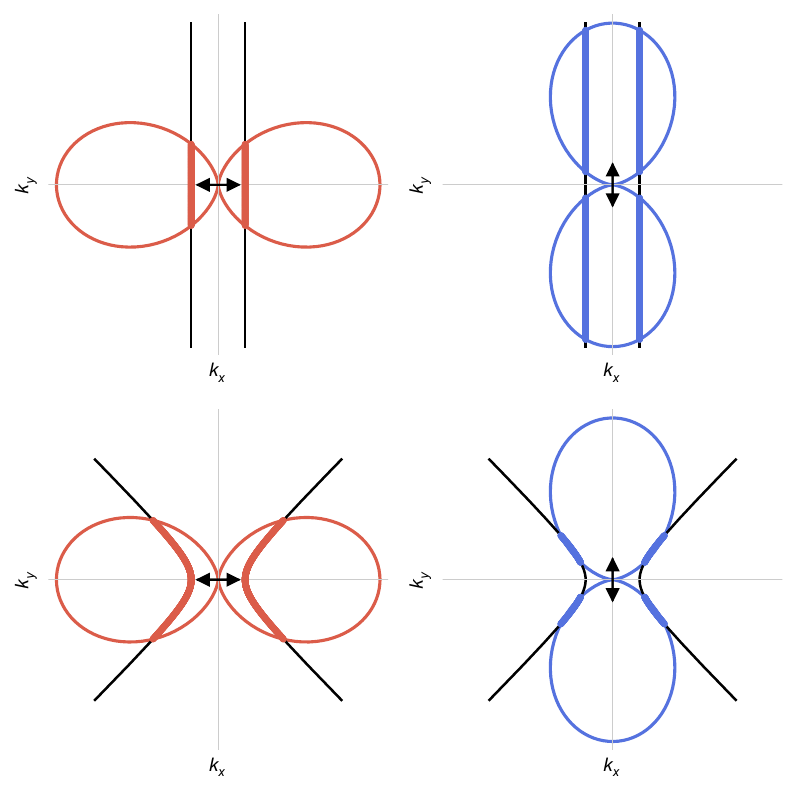}
    \caption{ \textbf{Modelling the rod direction as near-field dipole orientation.} Momentum-space overlap between dipole emission and polariton IFCs. The angular distribution of an in-plane dipole is superimposed on representative IFCs for canalized (top) and hyperbolic (bottom) regimes. For canalization, where the IFCs are nearly vertical, a dipole oriented along [010] provides stronger overlap with the IFCs, leading to more efficient excitation of polaritons propagating along [100]. In contrast, for hyperbolic dispersion with IFCs opening along $k_x$, a dipole oriented along [100] exhibits improved overlap. Colored segments highlight the portions of the IFCs lying within the dipole emission distribution, indicating regions of enhanced coupling.}
    \label{fig:pol_dipole_sketch}
\end{figure}

An intuitive understanding of the polarization dependence can be obtained from the overlap between the polariton isofrequency contours (IFCs) and the momentum distribution of an in-plane dipole representing the near-field excitation. The electric field of a dipole with moment $\mathbf{p}$ can be written as $\mathbf{E}(\mathbf{r}) \propto \mathbf{G}(\mathbf{r}) \cdot \mathbf{p}$, where $\mathbf{G}$ is the dyadic Green function $\mathbf{G} = (\mathbf{I} - \nabla\nabla/k_0^2) G_0$ with $G_0 = e^{ik_0 r}/(4\pi r)$ \cite{novotny2012principles}. In momentum space, the dyadic Green function can be expressed using the Weyl identity (angular spectrumrepresentation) as:
\begin{equation}
\mathbf{G}(\mathbf{k}) \propto 
\left( \mathbf{I} - \frac{\mathbf{k}\mathbf{k}}{k_0^2} \right) G_0(\mathbf{k}),
\end{equation}
where the scalar Green function in momentum space reads
\begin{equation}
G_0(\mathbf{k}) \propto \frac{e^{i k_z \lvert z_0\rvert}}{k_z},
\end{equation}
with $z_0$ the vertical position of the dipole and $k_z = \sqrt{k_0^2 - k_\parallel^2}$ the out-of-plane wavevector component. In the near-field regime relevant for polariton excitation, the dominant contributions arise from large in-plane momenta $k_\parallel \gg k_0$, for which $k_z \approx i k_\parallel$. In this limit, the second term of the Green tensor dominates, and the in-plane electric field becomes proportional to
\begin{equation}
\mathbf{E}(\mathbf{k}_\parallel) \propto \mathbf{G}(\mathbf{k}_\parallel) \cdot \mathbf{p}
\;\;\approx\;\;
\mathbf{k}_\parallel \cdot \mathbf{p},
\end{equation}
where $\mathbf{p}$ is the dipole moment. Therefore, the coupling strength at a given in-plane momentum is governed by the projection of the dipole onto the wavevector, and the corresponding emitted power scales as
\begin{equation}
\lvert\mathbf{E}(\mathbf{k}_\parallel)\rvert^2 \propto 
\lvert\mathbf{k}_\parallel \cdot \mathbf{p}\rvert^2
= k_\parallel^2 p^2 \cos^2\phi,
\end{equation}
where $\phi$ is the angle between the dipole moment $\mathbf{p}$ and the in-plane wavevector $\mathbf{k}_\parallel$. By superimposing the in-plane dipole emission onto the polariton isofrequency contours, one finds that a dipole oriented along the [010] direction provides a stronger overlap with the IFCs at the canalization condition (Fig. \ref{fig:pol_dipole_sketch}). In contrast, for hyperbolic dispersion where the IFCs open along the $k_x$ direction, a dipole oriented along [100] exhibits improved overlap, consistent with previous observations in MoOCl$_2$ at visible and near-infrared frequencies \cite{Venturi2024, li2025broadband}. A more rigorous treatment of the angular emission would require inclusion of substrate-induced modifications of the dipole field and the reflected part of the Green function~\cite{hu2023source}. However, the simplified picture presented here captures the essential physics underlying the enhanced excitation efficiency observed for polarization along [010] at the canalization wavelength.

\subsection{Experimental polariton launching under different polarization excitation}

\begin{figure}[h!]
    \centering
    \includegraphics[width=1\linewidth]{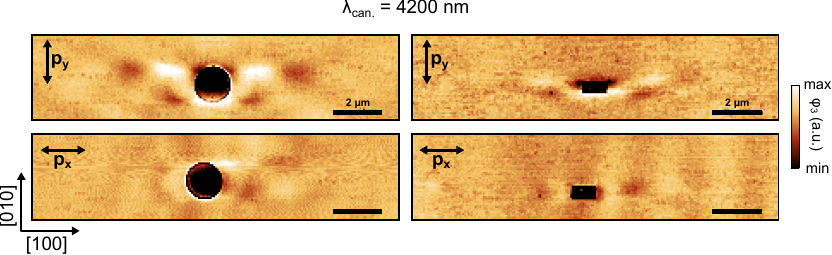}
    \caption{ \textbf{Near-field phase maps for different incident polarization directions.} sSNOM measurement performed at 4200~nm on a 45~nm-thick flake. A gold disk and a rod oriented along the [100] direction were used as plasmonic launchers (scale bars: 2 \textmu m). In both cases, when the polarization is along the [010], more efficient polariton launching is observed.}
    \label{fig:polarization_dependance_exp}
\end{figure}

To experimentally verify the change in the polariton launching efficiency with different polarizations, near-field s-SNOM measurements were performed while rotating the in-plane projection of the incident polarization along the two principal crystal axes (Fig.~\ref{fig:polarization_dependance_exp}). As expected, when the polarization is aligned along the [010] direction, the resulting maps display more pronounced and extended interference fringes, indicating more efficient polariton launching compared to excitation along the metallic [100] axis.

This behavior is consistently observed for both disk and rod launchers. The rod antenna is elongated along the [100] direction. Additionally, this configuration offers a practical advantage: excitation with polarization along [010] suppresses the generation of edge-launched polaritons from the flake boundaries. This effect can be observed in the rod-based measurements, where residual edge-launched contributions are present for the other polarization configuration. As a result, using polarization along [010] relaxes constraints on device geometry, avoiding the need for large flakes or positioning the launcher far from the edges to minimize unwanted contributions.

\newpage

\section{Extraction of experimental IFC curvature}\label{Exp_treat}

\begin{figure}[h!]
    \centering
    \includegraphics[width=1.0\linewidth]{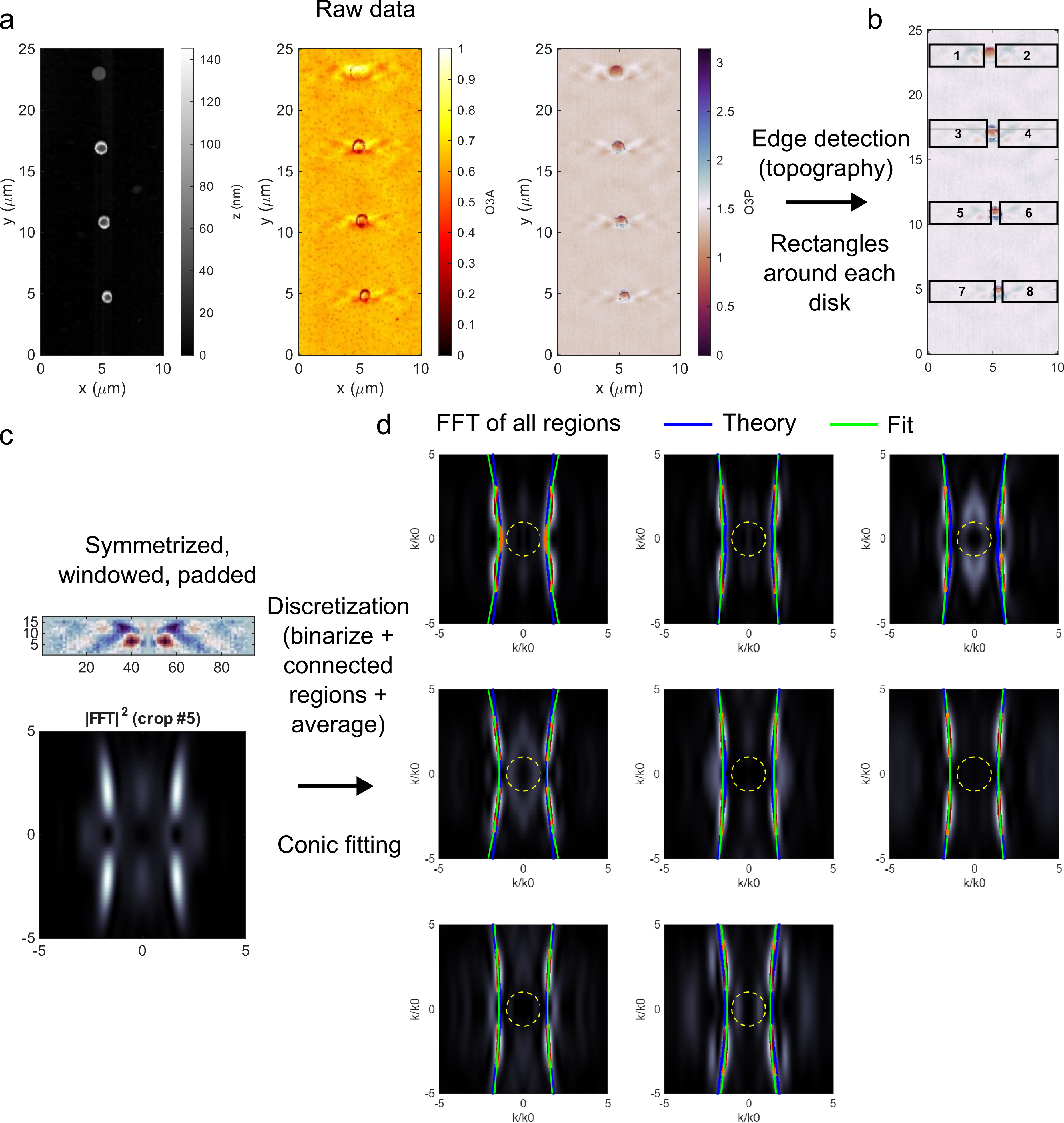}
    \caption{\textbf{Experimental data treatment for the k-space images of polaritons IFCs.} \textbf{a} Raw sSNOM data (topography, amplitude and phase) at $\lambda=\SI{3500}{\nm}$ for a $\SI{18}{\nm}$-thick flake. \textbf{b} Selection of separate regions to perform FFT obtained from edge detection of the disks from the topography image. \textbf{c} Before FFT we prepare the phase images by symmetrization and windowing in the center and at the edges in both directions. We then apply zero padding in order to increase the resolution of the FFT. \textbf{d} We discretize the FFT images by binarization and selecting the centerline of each connected region. A set of 2D points is extracted in this way, which we fit with a conic function, which allows the extraction of the IFCs curvature from the experiments.}
    \label{fig:Exp_data_treat}
\end{figure}

To quantitatively assess the topology of the experimental IFCs, we extract the vertex curvature (at $k_x = 0$) from the FFT maps. The full data-processing workflow is summarized in Fig.~\ref{fig:Exp_data_treat}. We begin with the raw sSNOM data—topography, amplitude, and phase maps. An example at $\lambda=\SI{3500}{\nm}$ for a $\SI{18}{\nm}$-thick flake is shown in Fig.~\ref{fig:Exp_data_treat}a. The launcher edges are extracted as the boundaries of the adequately thresholded topography maps. Using these edges, we define rectangular regions on the left and right sides of each disk (Fig.~\ref{fig:Exp_data_treat}b), and compute the FFT of each region separately. We use for the FFT computation only the phase data as they are generally less noisy than the amplitude ones and carry the same information for the purpose of the IFCs extraction. Prior to performing the FFT, we symmetrize the images about the axis where the scan intersects the disk. We additionally apply windowing on both the stitching boundary and the outer $x$ and $y$ edges, followed by zero-padding to increase the FFT resolution (Fig.~\ref{fig:Exp_data_treat}c). To extract the IFCs from the FFT amplitude maps, we discretize the resulting images as follows. We first binarize the FFT maps using an intensity threshold and then identify connected components using the \texttt{bwconncomp} function in MATLAB. We exclude from this procedure the central region indicated by the dashed yellow line where a strong DC component is sometimes present. Exploiting the expected four-lobe symmetry, we process each connected region row by row. For every row, the center point of a given region is determined via an intensity-weighted average of all pixels in that row belonging to the region. This yields a set of 2D points that trace the IFC (red points in Fig.~\ref{fig:Exp_data_treat}d, the color follows the intensity of the FFT map at that point). These point sets are then fitted to the conic form $k_x^2/a^2 + \mu\,k_y^2/b^2 = 1$, where $\mu=1$ corresponds to elliptical IFCs and $\mu=-1$ to hyperbolic IFCs. The fitted curves (green lines) accurately follow the theoretically calculated IFCs (blue lines). The vertex curvature is given by $\kappa=\mu\,a/b^2$. The final reported curvature is obtained by averaging the extracted values across all analyzed regions.

\newpage

\section{Experimental thickness-dependent canalization frequency}\label{Exp_canal_freq}

\begin{figure}[h!]
    \centering
    \includegraphics[width=1.0\linewidth]{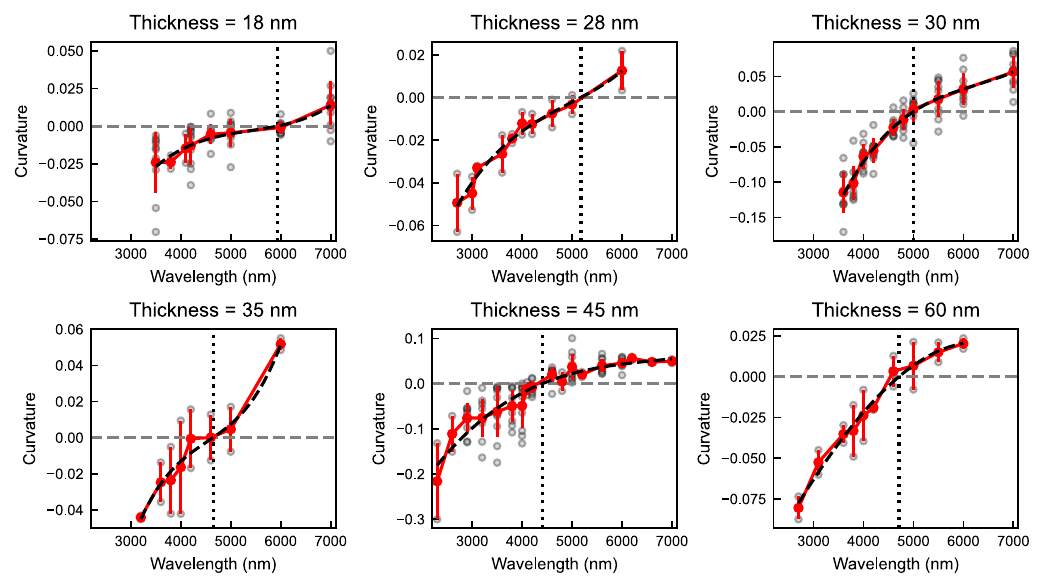}
    \caption{\textbf{Determination of the experimental thickness-dependent canalization frequency.} Fits of the experimental IFCs curvature against wavelength for all the MoOCl$_2$ flake thicknesses investigated in this work. Vertical dotted lines indicate the zero-crossing point of the IFCs corresponding to the canalization frequency.}
    \label{fig:all_thickness_fit}
\end{figure}

Given the extraction of the experimental IFCs and the corresponding curvature, we can determine the canalization frequency by finding the frequency at which the curvature changes sign. To increase the reliability of the analysis, we use the average curvature extracted from all the different regions of the images (both disk sides and for different disk diameters) as detailed in the previous section. To fully exploit the extracted dataset, we determine the zero-crossing point by fitting the experimental curvature versus wavelength dependence with a third-order polynomial. This phenomenological approach is adopted because the curvature cannot be expressed in closed analytical form as a function of wavelength, given that the IFCs are governed by an implicit dispersion relation (see eq. \eqref{eq:IFCs} below). A third-order polynomial was found to be the lowest-order function that consistently provides an accurate fit across all thicknesses. The fitted results for all the thicknesses investigated in this work are reported in Fig. \ref{fig:all_thickness_fit}. Gray dots are the single curvature values used to obtain the average (red dots), while black dashed curves are the polynomial fits. The spread of the individual data points mainly reflects variations associated with the excitation geometry and independent analysis of different launcher regions, although small local variations of the flake properties (e.g., residual strain, defects, or thickness inhomogeneities) may also contribute.

\newpage

\section{Theoretical thickness-dependent canalization frequency }\label{Theo_canal_freq}

\begin{figure}[ht]
    \centering
    \includegraphics[width=0.75\linewidth]{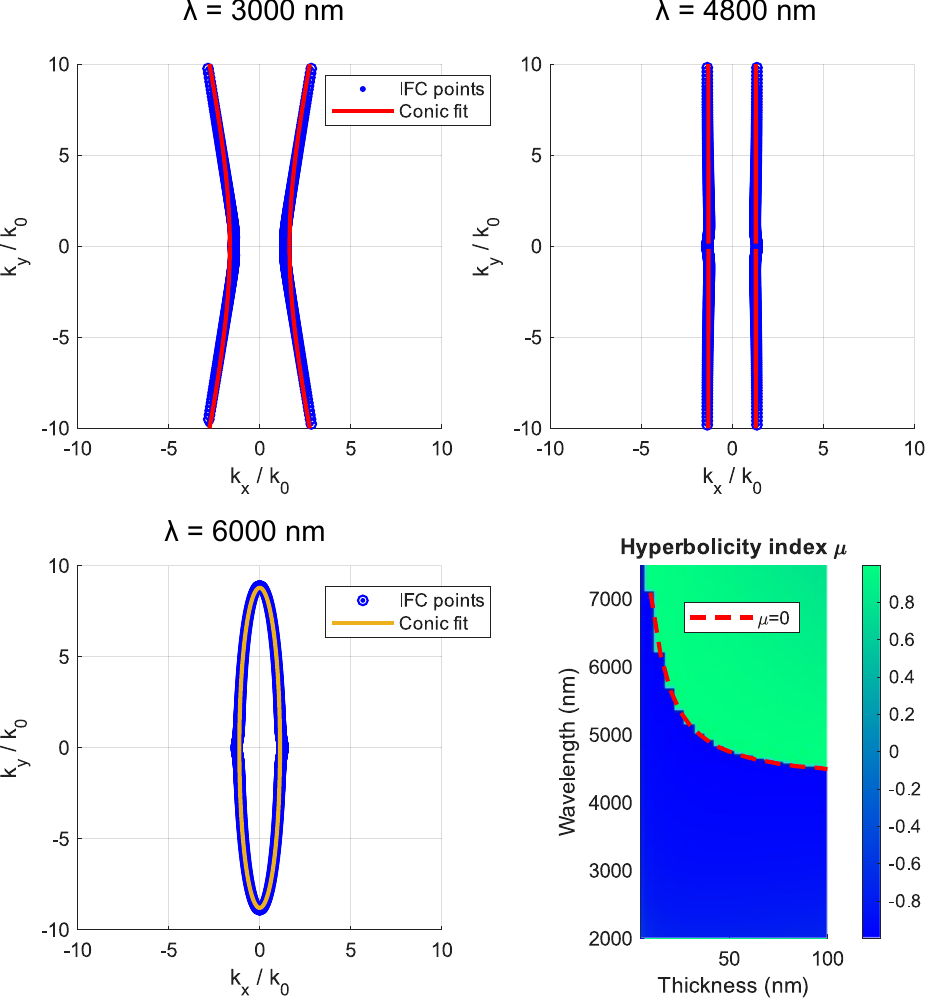}
    \caption{\textbf{Determination of the theoretical thickness-dependent canalization frequency.} Calculated IFCs for a 45 nm film thickness at various frequencies. The IFCs are fitted with the conic function. The color of the fit changes depending on the value of $\mu$ determining the IFCs topology. We also show the value of $\mu$ across the wavelength and thickness reported in Fig. 3 of the main text.}
    \label{fig:th_canal_shift}
\end{figure}

\begin{figure}[ht]
    \centering
    \includegraphics[width=1\linewidth]{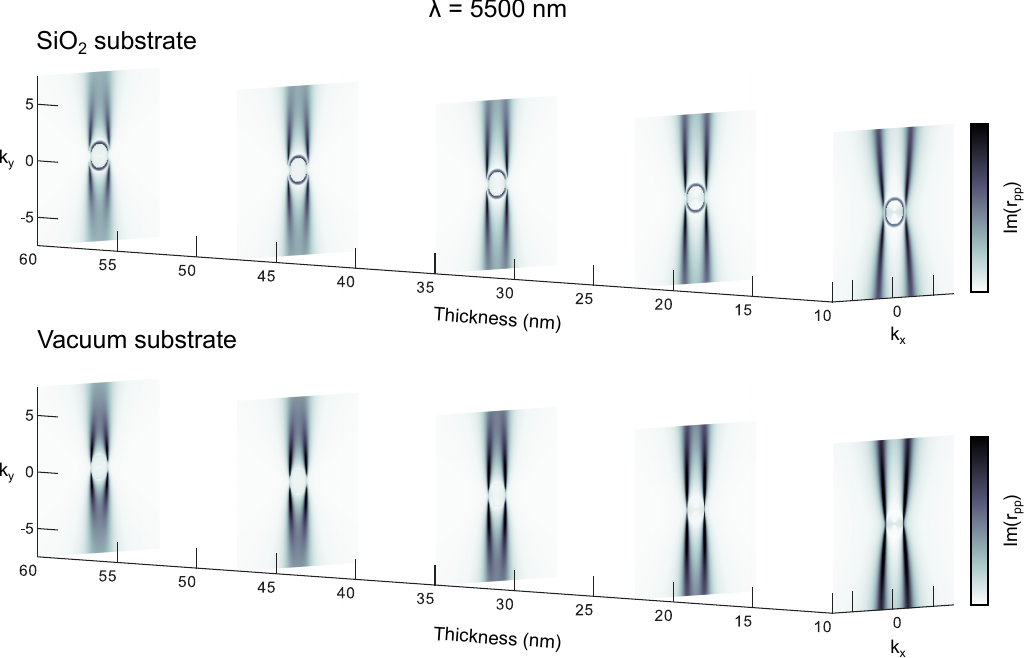}
    \caption{\textbf{Effect of substrate refractive index and dielectric function on canalization frequency shift.} Transfer matrix calculation of the evolution of the IFCs with film thickness for a SiO$_2$ substrate (top) or a vacuum substrate (bottom).}
    \label{fig:shift_intuition}
\end{figure}

In the limit of small thickness compared to the free-space wavelength $k_0d \ll 1$, the in-plane wavevector of polaritons in a biaxial media can be obtained as the solution of the following equation \cite{alvarez2019analytical, Venturi2024}:

\begin{equation}\label{eq:IFCs}
\begin{aligned}
&\Bigg[
\frac{k_0 d \,\varepsilon_x}{2 i}\, k_y^2
+ \frac{k_0 d \,\varepsilon_y}{2 i}\, k_x^2
+ \frac{k_x^2 + k_y^2}{2}
\left(
i \sqrt{k_x^2 + k_y^2 - \varepsilon_1}
+ i \sqrt{k_x^2 + k_y^2 - \varepsilon_3}
\right)
\Bigg] \\
&\times
\Bigg[
\frac{k_0 d \,\varepsilon_x}{2 i}\, k_x^2
+ \frac{k_0 d \,\varepsilon_y}{2 i}\, k_y^2
+ \frac{k_x^2 + k_y^2}{2}
\left(
\frac{\varepsilon_1}{i \sqrt{k_x^2 + k_y^2 - \varepsilon_1}}
+ \frac{\varepsilon_3}{i \sqrt{k_x^2 + k_y^2 - \varepsilon_3}}
\right)
\Bigg] \\
&= - \frac{k_0^2 d^2}{4}\,
k_x^2 k_y^2 \,
\left(\varepsilon_x - \varepsilon_y\right)^2
\end{aligned}
\end{equation}

where $\varepsilon_1$ and $\varepsilon_3$ are the permittivities of the media above and below the biaxial film of thickness $d$. From the numerical solution of this equation, we obtain the theoretical IFCs shown in Fig. 1 and Fig. 2 of the main text. From the analytical IFCs we can also predict the change in canalization frequency with film thickness. We perform a similar procedure to what we apply to the experimental data: we fit the analytical IFCs with the conic equation and extract the vertex curvature. In this way we construct the image shown in Fig. 3e of the main text.
Examples of the analytical IFCs and corresponding conic fits with equation $k_x^2/a^2 + \mu\,k_y^2/b^2 = 1$ are displayed in Fig. \ref{fig:th_canal_shift} for a \SI{45}{\nano \meter} thick film. We also report the extracted values of $\mu$ from the conic fit, which highlight the switch from the hyperbolic $\mu=-1$ to the elliptical regime $\mu=1$.

\begin{figure}[ht]
    \centering
    \includegraphics[width=1\linewidth]{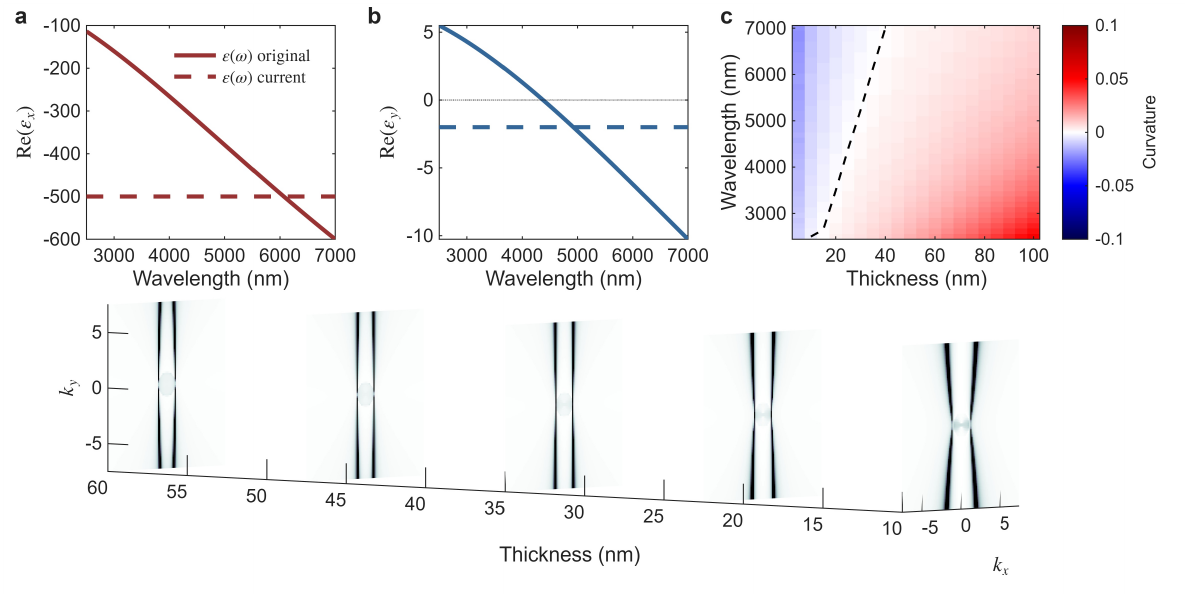}
    \caption{\textbf{Canalization shift for a material with constant dielectric function.} The used permittivity (dashed lines, $\varepsilon_x=-500+100i$ and $\varepsilon_y=-2+0.1i$) is shown along the MoOCl$_2$ mid IR dielectric function along the \textbf{a} x and \textbf{b} y directions. \textbf{c} Shift of the canalization frequency computed from the analytical dispersion as inferred from the zero of the IFC vertex curvature. \textbf{d} IFCs calculated from the transfer matrix at \SI{5500}{\nano \meter} for flakes of different thicknesses.}
    \label{fig:canshift_xcnst_ycnst}
\end{figure}

The theoretical dispersion can also be used to gather an intuitive explanation of the redshift of the canalization frequency for decreasing flake thickness. First of all, we make sure that the presence of the substrate does not strongly affect the observed frequency shift (Fig. \ref{fig:shift_intuition}a). This is confirmed as we observe an almost identical thickness-evolution of the IFCs for a SiO$_2$ or vacuum substrates (at $\lambda = \SI{5500}{\nano \meter}$). The only noticeable difference is the appearance of a circular component at the light line of the substrate in the SiO$_2$ case. Indeed, for substrates with a higher refractive index, the IFC shape can be strongly modified as they cannot enter the substrate light line and are thus pushed to higher wavevectors in a non-uniform way (they follow the circular shape of the light line). 

\begin{figure}[ht]
    \centering
    \includegraphics[width=1\linewidth]{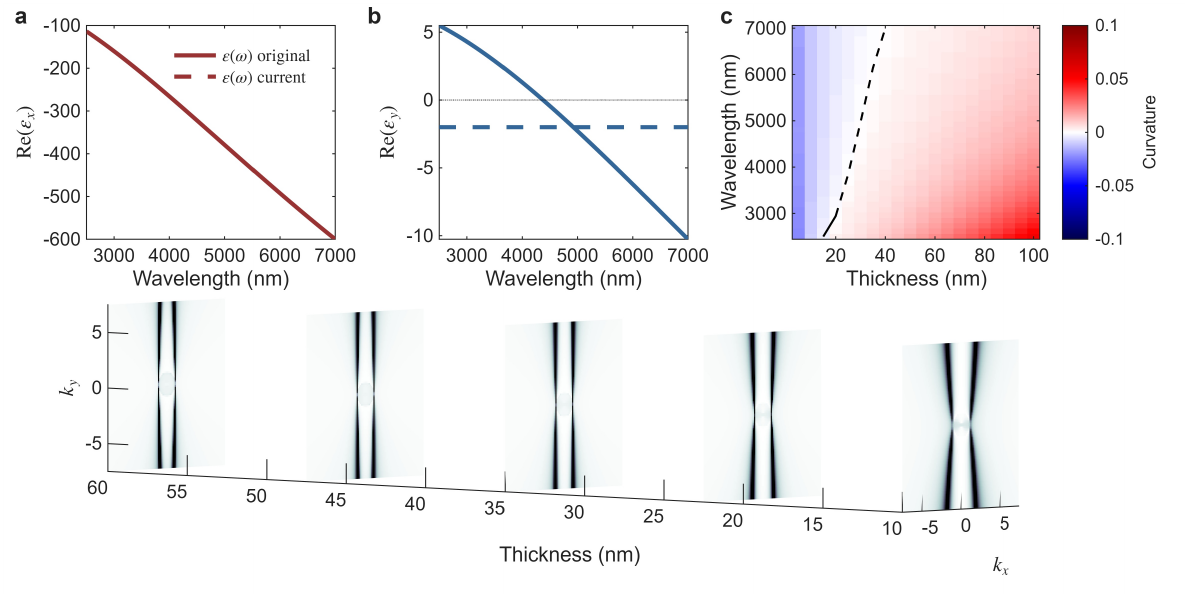}
    \caption{\textbf{Canalization shift for a material with the mid-IR dielectric function of MoOCl2 along x and a constant permittivity along y .} The used permittivity (dashed lines, $\varepsilon_y=-2+0.1i$) is shown along the MoOCl$_2$ mid IR dielectric function along the \textbf{a} x and \textbf{b} y directions. \textbf{c} Shift of the canalization frequency computed from the analytical dispersion as inferred from the zero of the IFC vertex curvature. \textbf{d} IFCs calculated from the transfer matrix at \SI{5500}{\nano \meter} for flakes of different thicknesses.}
    \label{fig:canshift_xMoOCl2_ycnst}
\end{figure}

To further clarify the physical origin of the canalization wavelength shift with thickness, we first consider the simplest possible case of a material with a non-dispersive dielectric response ($\varepsilon_x=-500+100i$ and $\varepsilon_y=-2+0.1i$, Fig. \ref{fig:canshift_xMoOCl2_ycnst}a, b), chosen such that canalization occurs at $\lambda = 5500$~nm for a thickness of 35~nm. In this analysis we restrict ourselves to a layer suspended in vacuum in order to exclude any contribution from the substrate refractive index. The corresponding IFC evolution, calculated via transfer-matrix methods, is shown in Fig. \ref{fig:canshift_xMoOCl2_ycnst}d. Interestingly, even for a completely frequency-independent dielectric function, a clear transition from hyperbolic to elliptical IFCs is observed as the thickness is varied. This demonstrates that the thickness-dependent shift of the topological transition is not solely driven by the dispersion of the dielectric function, but also by the intrinsic scaling between the free space momentum and the film thickness.

This behavior can be understood directly from eq. \eqref{eq:IFCs}, where the thickness always appears multiplied by the free-space wavevector $dk_0$. As a consequence, increasing the thickness by a factor $x$ can be compensated by decreasing the free-space wavevector by the same factor, corresponding to a linear increase of the canalization wavelength with thickness. Analytical calculations of the IFC evolution with tracking of the canalization wavelength shift with film thickness confirm this trend for the non-dispersive dielectric function (Fig. \ref{fig:canshift_xMoOCl2_ycnst}c).

This behavior is opposite to what is experimentally observed in MoOCl$_2$, where increasing the thickness produces a redshift of the canalization wavelength. Therefore, the dispersion of the Drude dielectric function must not only compensate but invert the intrinsic geometrical scaling described above. Since the dielectric-dependent terms enter the dispersion in the form $d,\varepsilon(\lambda_0)/\lambda_0$, a dielectric function whose magnitude decreases at longer wavelengths naturally reverses the trend imposed by the geometrical scaling, producing the experimentally observed evolution of the canalization condition.

To further isolate the role of the dielectric dispersion along the two principal axes, we separately analyze the effect of allowing only one component of the permittivity tensor to remain dispersive. We first consider the case where the $y$-direction is kept constant ($\varepsilon_y=-2+0.1i$), while the $x$-direction follows the dispersive dielectric function of MoOCl$_2$ (Fig.~\ref{fig:canshift_xMoOCl2_ycnst}a, b). The resulting IFC evolution (Fig.~\ref{fig:canshift_xMoOCl2_ycnst}d) is found to be very similar to the case of fully constant dielectric function, with the canalization wavelength still scaling approximately linearly with thickness (Fig.~\ref{fig:canshift_xMoOCl2_ycnst}c). This indicates that the dispersion along the $x$ direction has only a minor influence on the canalization frequency shift. Such behavior is expected, as $\varepsilon_x$ remains strongly negative over the investigated spectral range, making the material response along this direction close to that of a perfect electric conductor.

\begin{figure}[ht]
    \centering
    \includegraphics[width=1\linewidth]{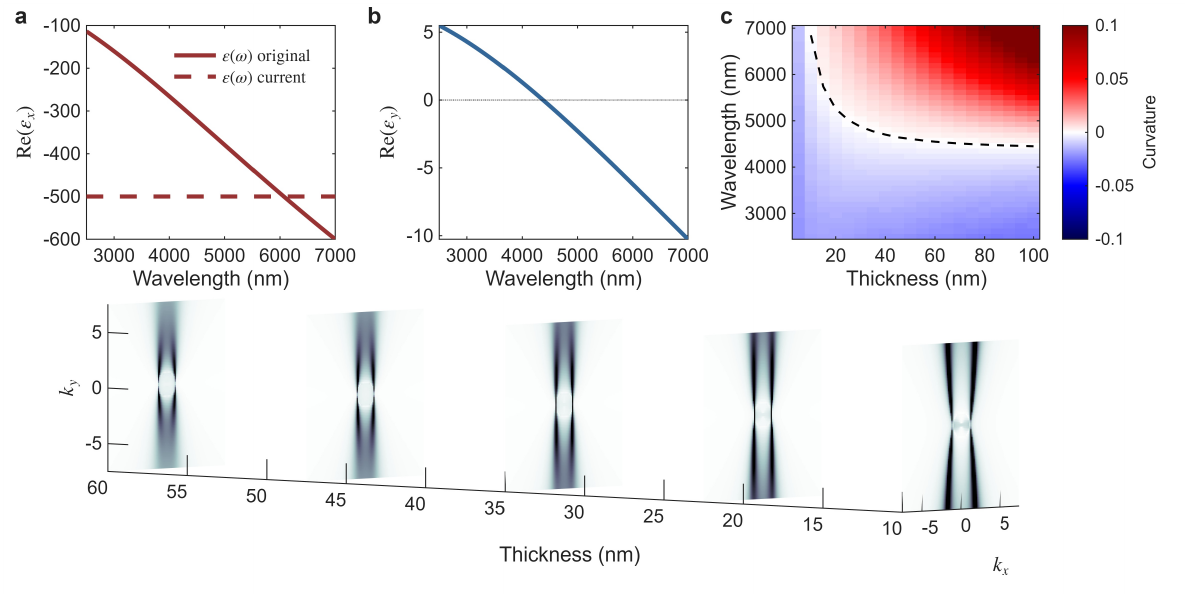}
    \caption{\textbf{Canalization shift for a material with the mid-IR dielectric function of MoOCl2 along y and a constant permittivity along x.} The used permittivity (dashed lines, $\varepsilon_x=-500+100i$) is shown along the MoOCl$_2$ mid IR dielectric function along the \textbf{a} x and \textbf{b} y directions. \textbf{c} Shift of the canalization frequency computed from the analytical dispersion as inferred from the zero of the IFC vertex curvature. \textbf{d} IFCs calculated from the transfer matrix at \SI{5500}{\nano \meter} for flakes of different thicknesses.}
    \label{fig:canshift_xcnst_yMoOCl2}
\end{figure}

We then consider the opposite limit, where $\varepsilon_x=-500+100i$ is kept constant while the dielectric response along the $y$ direction follows the dispersive permittivity of MoOCl$_2$ in the mid-infrared (Fig.~\ref{fig:canshift_xcnst_yMoOCl2}a, b). In this case, the IFCs evolve much more rapidly toward the elliptical regime as the thickness increases (Fig.~\ref{fig:canshift_xcnst_yMoOCl2}d), leading to a qualitatively different behavior from the previous cases. Tracking the canalization wavelength through the analytical IFC calculations reveals a trend that resembles the experimentally observed evolution (Fig.~\ref{fig:canshift_xcnst_yMoOCl2}c). These results demonstrate that the shift of the canalization wavelength is predominantly governed by the dispersion of the dielectric function along the $y$ direction, while the dispersion along $x$ only introduces minor quantitative corrections.

\begin{figure}[ht]
    \centering
    \includegraphics[width=1\linewidth]{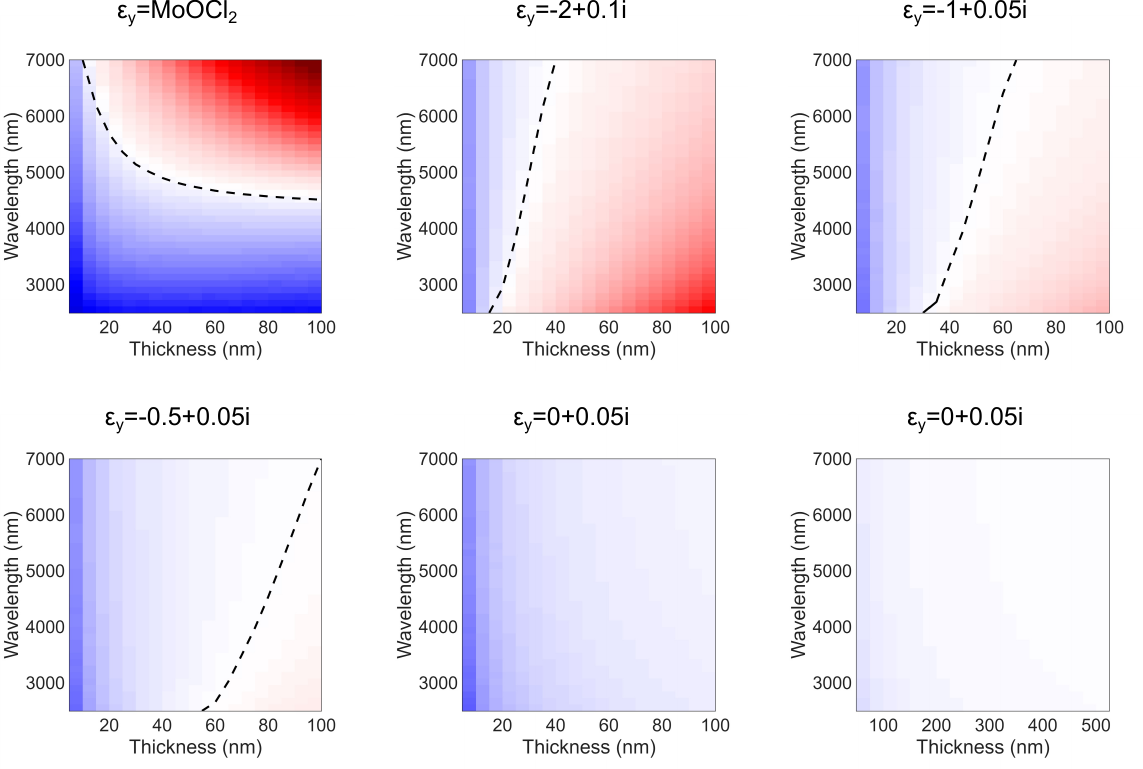}
    \caption{\textbf{Canalization shift for materials with the mid-IR dielectric function of MoOCl2 along x and a near-zero constant permittivity along y.} Calculated IFCs curvature in the small-thickness limit for different constant values of the [010] permittivity. Dashed black lines indicate the canalization condition.}
    \label{fig:canshift_xcnst_ycnst_evo}
\end{figure}

To assess the role of the [010] permittivity in setting the canalization condition, we calculated the IFC curvature evolution while keeping $\varepsilon_x(\omega)$ equal to the fitted MoOCl$_2$ response and fixing $\varepsilon_y$ to selected non-dispersive values around zero (Fig. \ref{fig:canshift_xcnst_ycnst_evo}). This allows us to isolate the effect of the [010] dielectric response independently of its dispersion.

For negative values of $\varepsilon_y$, the calculated zero-curvature line remains linear as a function of thickness, with values closer to zero shifting the canalization condition toward larger thicknesses and longer wavelengths. In contrast, when $\varepsilon_y$ becomes positive (here we analyse the $Re(\varepsilon_y)=0$, as the same happens for small positive values of $Re(\varepsilon_y)$), the IFCs remain hyperbolic within the thin-film approximation and the curvature approaches zero only asymptotically, without crossing into the elliptical regime. This behavior is also observed when extending the calculation to a broader thickness range, where the curvature remains small and negative over a wide spectral region (last panel of Fig. \ref{fig:canshift_xcnst_ycnst_evo}).

These results show that a fixed small positive dielectric response along [010] is not sufficient to obtain a true hyperbolic-to-elliptical topological transition. We note that this analysis is performed within the thin-film approximation. At larger thicknesses, this approximation is expected to gradually break down, and the out-of-plane response $\varepsilon_z$ also plays a role in determining the IFC shape. In this regime, the canalization condition is no longer governed solely by the in-plane components $\varepsilon_x$ and $\varepsilon_y$, and the topological transition depends on the full biaxial dielectric tensor and its analysis more complicated.

\subsection{Canalization wavelength shift in a Lorentz-Lorentz material}

To assess whether thickness-dependent tuning of the canalization wavelength is expected in other van der Waals systems, we considered a representative Lorentz--Lorentz dielectric response corresponding to the typical case of phonon-polariton materials with Reststrahlen bands along both in-plane axes. As a model system, we started from the in-plane dielectric function of hBN \cite{ambrosio2018selective} and rescaled the phonon frequencies such that the longitudinal optical phonon, and therefore the relevant sign-changing region of the permittivity, lies near 4.2~$\mu$m, comparable to the intrinsic canalization wavelength of MoOCl$_2$. The relative oscillator strength was preserved by keeping the quantity $(\omega_{\mathrm{LO}}^2-\omega_{\mathrm{TO}}^2)/\omega_{\mathrm{TO}}^2$ constant, ensuring that the spectral rescaling does not artificially modify the dipole strength of the phonon resonance. An in-plane anisotropy was then introduced by shifting the second Reststrahlen band by a constant factor, applying the same scaling to $\omega_{\mathrm{TO}}$, $\omega_{\mathrm{LO}}$, and the damping rate. This generates adjacent hyperbolic and elliptical regions separated by a natural topological transition.

\begin{figure}[ht]
    \centering
    \includegraphics[width=1\linewidth]{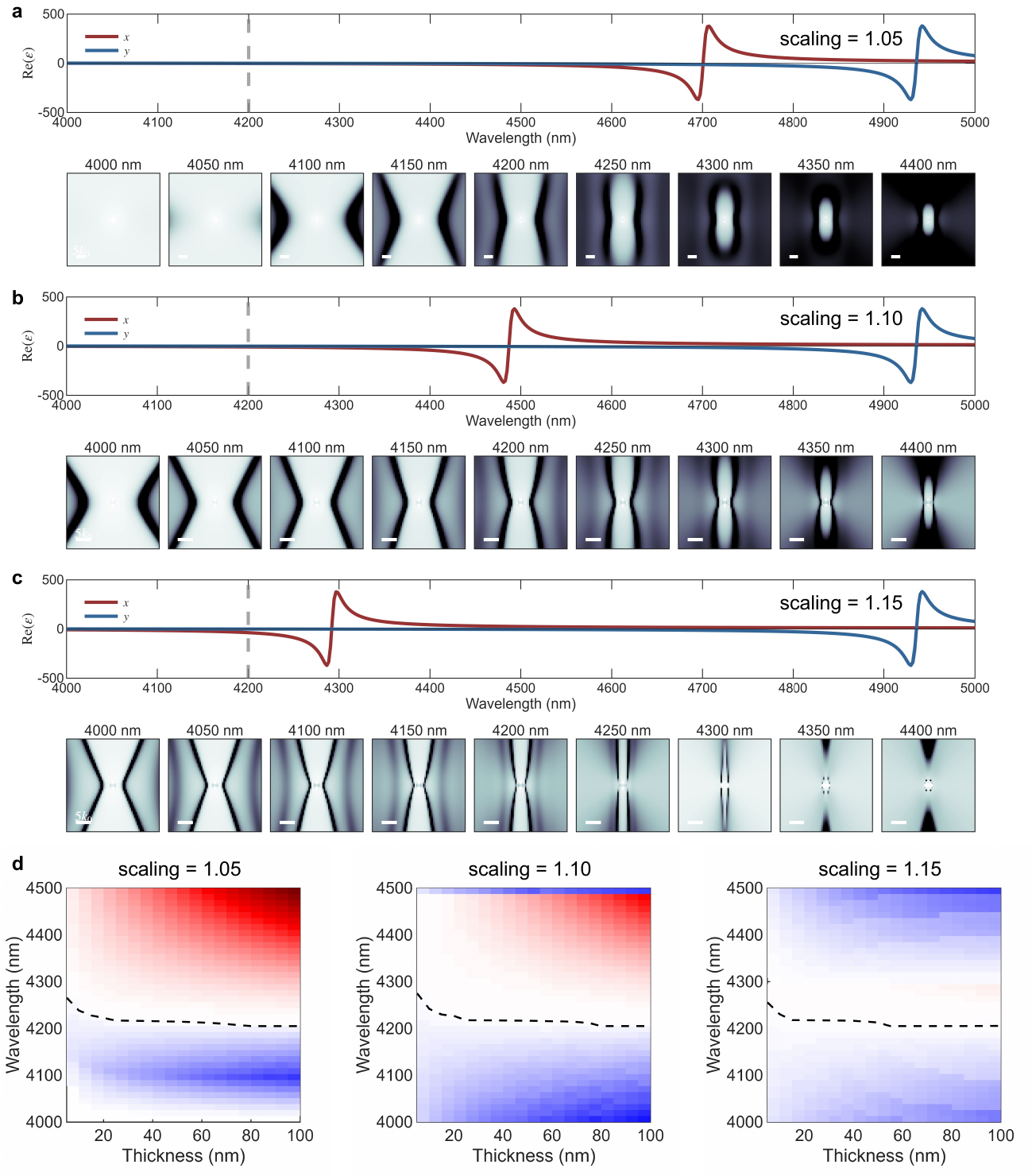}
    \caption{\textbf{Canalization wavelength shift in a model Lorentz-Lorentz material.} Real part of the in-plane dielectric function, with the dashed gray line marking the 4200~nm wavelength where the $y$-axis permittivity crosses zero. Each permittivity is characterized by the scaling between the two Lorentz oscillators with values of \textbf{a}~1.05, \textbf{b}~1.10 and \textbf{c}~1.15. For each permittivity, the corresponding isofrequency contours calculated via transfer-matrix methods in the 4000-4500~nm spectral range are shown. \textbf{d}~Extracted canalization wavelength as a function of thickness, obtained from analytical IFC calculations and conic fitting of the IFC curvature for all three permittivities.}
    \label{fig:canshift_Lor_Lor}
\end{figure}

The resulting dielectric functions and corresponding IFCs evolution are shown in Fig. \ref{fig:canshift_Lor_Lor} for three anisotropy values (scaling = 1.05, 1.10, and 1.15 in panels a, b and c respectively). The upper panels report the real part of the dielectric function along the two in-plane axes, with the dashed gray line marking the wavelength of 4200~nm corresponding to the zero-crossing of the $y$-axis permittivity (around which the canalizaiton is expected to be). The permittivity sign change in the $x$ direction is shifted toward shorter wavelengths, reproducing a situation analogous to MoOCl$_2$ where the two axes undergo the sign transition at different positions.

The middle panels show the corresponding isofrequency contours calculated via the transfer-matrix method in the 4400-4000~nm spectral range for a representative 50~nm film. As the separation between the two Reststrahlen bands increases, the confinement of the modes decreases due to the increasingly negative value of $\varepsilon_x$, accompanied by reduced losses (sharper IFCs) resulting from weaker field penetration into the material. In all cases, the intrinsic topological transition between hyperbolic and elliptical propagation appears in the narrow 4300-4200~nm spectral range.

Fig. \ref{fig:canshift_Lor_Lor}d tracks the evolution of the canalization wavelength with thickness for the three anisotropy values, obtained from analytical IFC calculations and extraction of the curvature through the same conic fitting procedure employed in the main text. In all cases, the canalization wavelength exhibits only a weak dependence on thickness and remains confined within the narrow 4300-4200~nm range. This behavior contrasts strongly with MoOCl$_2$, where over the same thickness range the canalization wavelength can be tuned by more than 1~$\mu$m. These results highlight how a Drude-Drude permittivity is substantially more favorable than a Lorentz-Lorentz one for achieving intrinsically broadband and thickness-tunable canalization.

\clearpage

\newpage

\section{Additional sSNOM phase maps}\label{SNOM_maps}

\begin{figure}[h!t]
    \centering
    \includegraphics[width=1.0\linewidth]{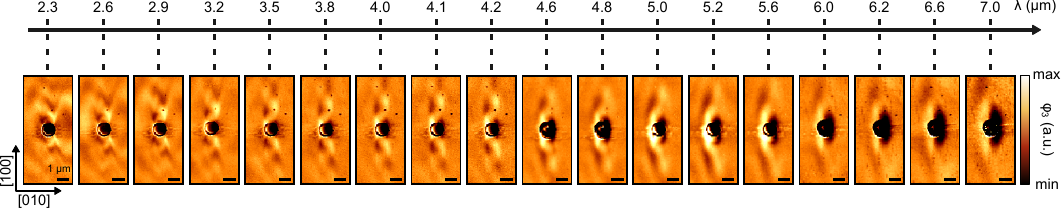}
    \caption{ \textbf{Near-field imaging of the wavelength-driven transition with a disk launcher.} A gold disk is used as the plasmonic launcher while the excitation wavelength is swept across the hyperbolic, canalization, and elliptic regimes (scale bars: 1 \textmu m, flake thickness \SI{45}{\nano \meter}).}
    \label{fig:Wavelength_sweep_disk}
\end{figure}

Figure~\ref{fig:Wavelength_sweep_disk} presents near-field phase maps of the third demodulation order obtained on a 45~nm-thick $\mathrm{MoOCl}_{2}$ flake using a gold disk as the plasmonic launcher while sweeping the excitation wavelength. The polarization of the incident light is aligned along the dielectric [010] axis of the flake to maximize the excitation efficiency of the plasmon-polariton modes. At shorter wavelengths, the near-field phase maps display strongly diverging hyperbolic wavefronts, characteristic of the hyperbolic dispersion regime, where the in-plane permittivity components have opposite signs. As the wavelength increases, the wavefronts gradually straighten, evolving into a highly directional, non-diverging pattern that signifies the onset of the canalization regime, where optical energy propagates in a nearly diffraction-free manner. Upon further increasing the wavelength, the wavefronts bend and form closed, elliptical shapes, corresponding to the elliptic dispersion regime, where both in-plane permittivity components are of the same sign.

\begin{figure}[h!t]
    \centering
    \includegraphics[width=1.0\linewidth]{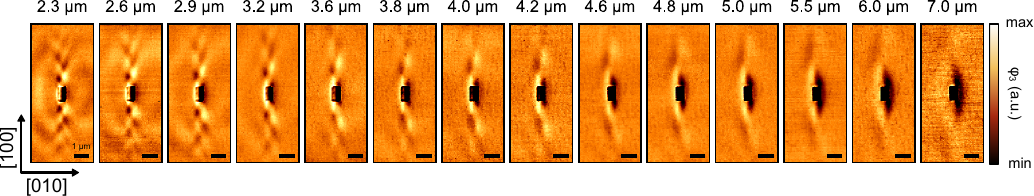}
    \caption{ \textbf{Near-field imaging of the wavelength-driven transition with a rod launcher.} A gold nanorod is used as the plasmonic launcher while the excitation wavelength is swept across the hyperbolic, canalization, and elliptic regimes (scale bars: 1 \textmu m, flake thickness \SI{45}{\nano \meter}).}
    \label{fig:Wavelength_sweep_rods}
\end{figure}

This continuous evolution of wavefronts provides a direct visualization of the topological transition from hyperbolic to elliptic propagation, with the canalization regime occurring at intermediate wavelengths. The observation highlights the intrinsic tunability of plasmon-polariton modes in $\mathrm{MoOCl}_{2}$ through wavelength variation alone. Importantly, the same qualitative behavior is observed when using a gold nanorod launcher (Fig.~\ref{fig:Wavelength_sweep_rods}), demonstrating that the mode evolution and the emergence of canalization are robust features of the material and do not strongly depend on the specific geometry of the plasmonic launcher (at least in the subwavelength limit here investigates). These results collectively confirm that $\mathrm{MoOCl}_{2}$ flakes support broadband, room-temperature plasmon-polariton canalization, with a mode transition that can be continuously tuned by the excitation wavelength.

\begin{figure}[h!]
    \centering
    \includegraphics[width=1.0\linewidth]{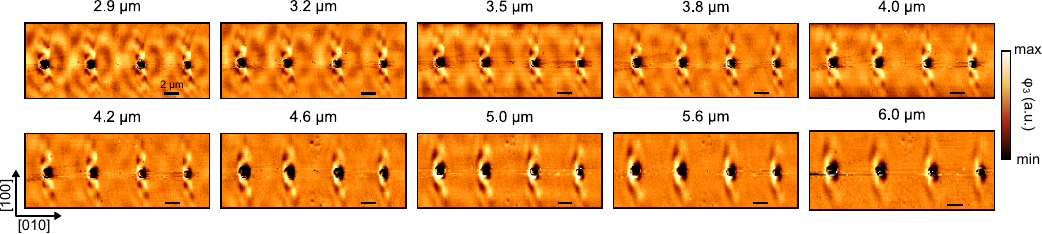}
    \caption{\textbf{Near-field imaging of the wavelength-driven transition with 4 disks launchers.} Gold disks of different sizes are used as the plasmonic launchers while the excitation wavelength is swept across the hyperbolic, canalization, and elliptic regimes (scale bars: 2 \textmu m, flake thickness \SI{45}{\nano \meter}).}
    \label{fig:Wavelength_sweep_four_disks}
\end{figure}

\begin{figure}[h!t]
    \centering
    \includegraphics[width=1.0\linewidth]{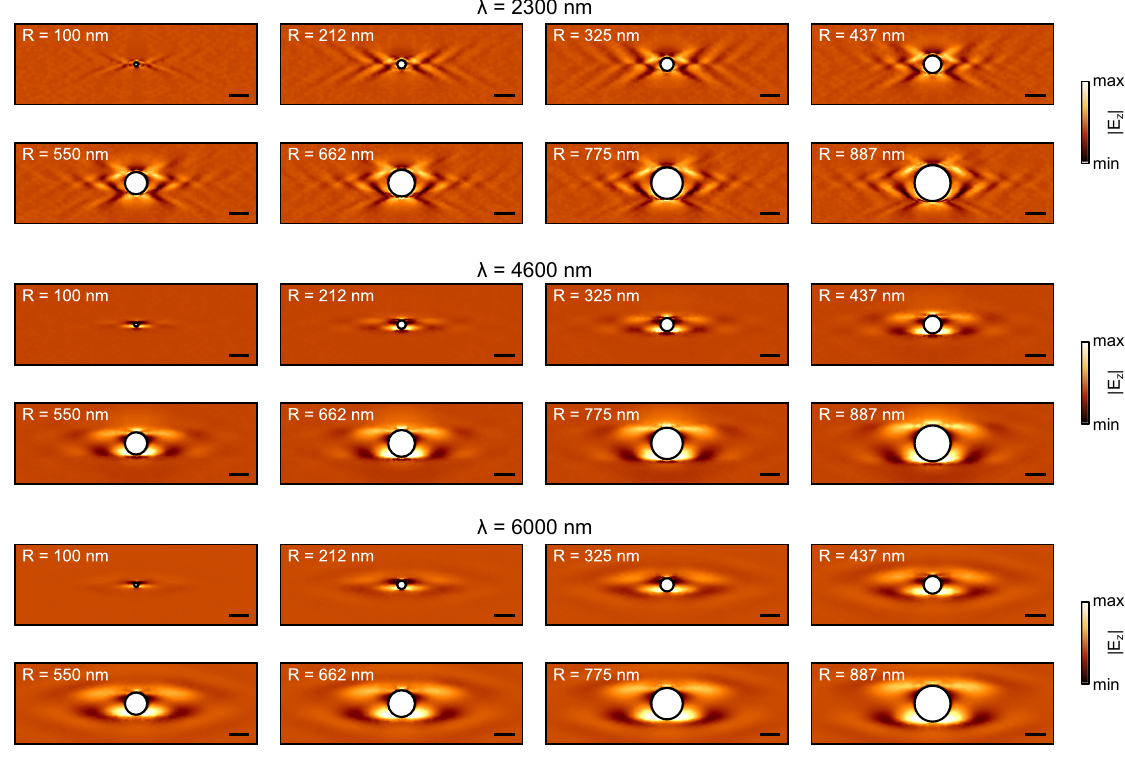}
    \caption{\textbf{Simulated out-of-plane electric field amplitude maps showing the wavelength-driven transition with a disk launcher.} A gold disk is used as the plasmonic launcher, with the disk radius varied from \SI{100}{\nano\meter} to \SI{887}{\nano\meter}. The flake thickness in the simulation is \SI{45}{\nano\meter}. Simulations were performed for three representative wavelengths corresponding to the hyperbolic, canalization, and elliptical propagation regimes (scale bars: 1 \textmu m). }
    \label{fig:SI_sweep_disk_size}
\end{figure}

The size of the gold launcher does not significantly affect the overall launching efficiency or the type of plasmonic modes excited (Fig.~\ref{fig:Wavelength_sweep_four_disks}). However, simulations (Fig.~\ref{fig:SI_sweep_disk_size}) reveal subtle differences in the spatial distribution of the launched waves depending on the disk radius. For small, point-like disks (R = 100~nm), the launcher behaves effectively as a single point source, producing a concentrated hot spot from which polaritons propagate along the x-axis (metallic [100] crystal direction). As the disk radius increases, the induced oscillating charges driven by the incident polarization along the y-axis ([010] direction) are no longer uniform across the disk surface. This results in stronger near-field amplitudes at the top and bottom edges along the polarization axis, creating two effective excitation regions that act as separate sources for polaritons propagating along x. The appearance of these multiple hot spots can be understood as a combination of geometric extension of the launcher, anisotropic coupling efficiency of the disk edges to the polariton modes, and interference effects between the induced near-field regions. For even larger disks, additional portions of the disk contribute weaker fields, producing more complex near-field patterns and slight distortions of the propagating wavefronts, yet the fundamental mode type and propagation direction remain unchanged. These observations indicate that the multiple launching spots are not indicative of different mode types but rather arise from the spatial distribution of induced charges and the anisotropic coupling between the launcher and the plasmon-polariton mode, as confirmed experimentally (Fig.~\ref{fig:Wavelength_sweep_four_disks}). This demonstrates that the plasmonic mode structure is primarily governed by the intrinsic anisotropic properties of $\mathrm{MoOCl}_{2}$, while the launcher geometry mainly modulates the spatial profile of the excitation. 

\begin{figure}[h!t]
    \centering
    \includegraphics[width=1.0\linewidth]{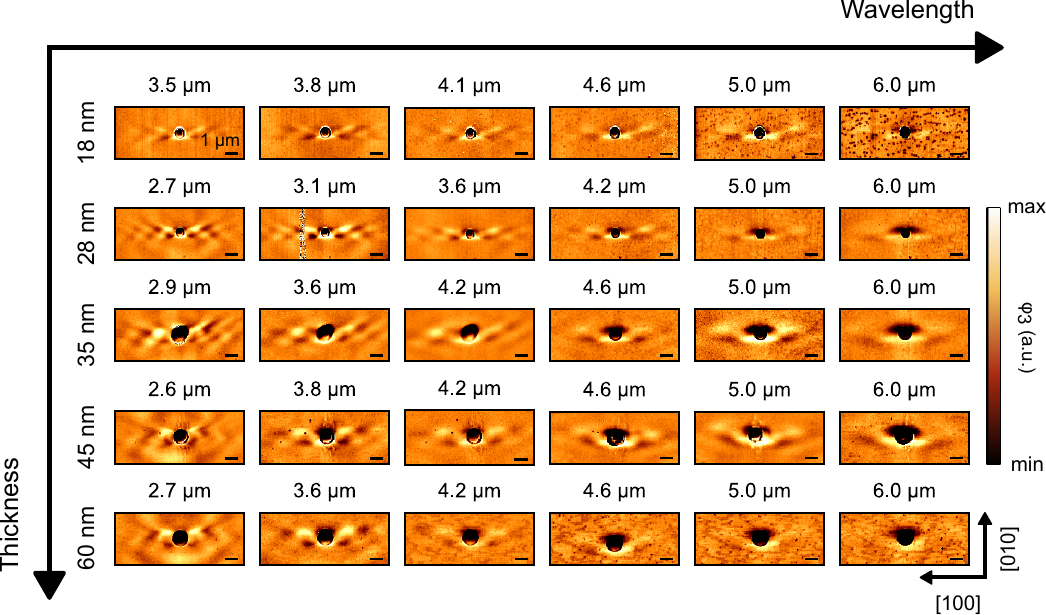}
    \caption{\textbf{Thickness–wavelength dependent matrix.} Near-field phase maps acquired by sSNOM with the excitation polarization aligned along [010], showing the dependence of plasmon-polariton canalization on incident wavelength and flake thickness.}
    \label{fig:thickness_wl_matrix}
\end{figure}

To illustrate the transition of the plasmon-polariton modes for different flake thicknesses, we plotted a thickness–wavelength dependent matrix (Fig.~\ref{fig:thickness_wl_matrix}). This matrix provides a comprehensive visualization of how the wavefronts evolve across the hyperbolic, canalization, and elliptical regimes as a function of both excitation wavelength and flake thickness, clearly showing how the canalization wavelength shifts toward shorter values for increasing flake thickness. It highlights the tunability of the canalization effect through an intrinsic material parameter. 

\newpage

\section{Propagation length of anisotropic plasmon polaritons}\label{Propagation_length}

\begin{figure}[h!t]
    \centering
    \includegraphics[width=1.0\linewidth]{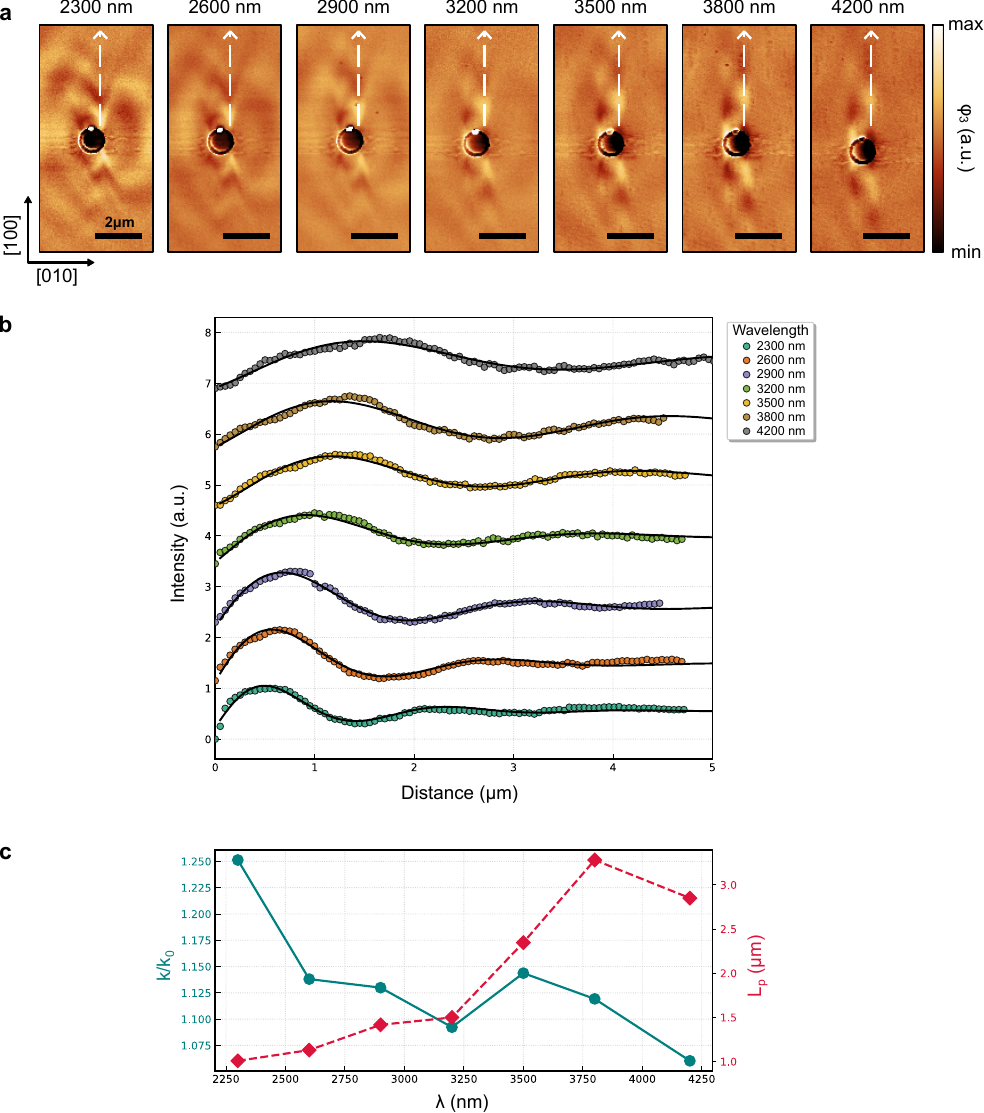}
    \caption{\textbf{Polariton propagation length on a \SI{45}{\nano \meter} flake launched by a disk antenna.} \textbf{a} sSNOM phase maps from which decay profiles are extracted along the white dashed lines. \textbf{b} Extracted profiles fitted with an exponentially decaying function. Curves are offset for clarity. \textbf{c} Retrieved wavevector and propagation length as a function of wavelength.}
    \label{fig:prop_disk}
\end{figure}

\begin{figure}[h!t]
    \centering
    \includegraphics[width=1.0\linewidth]{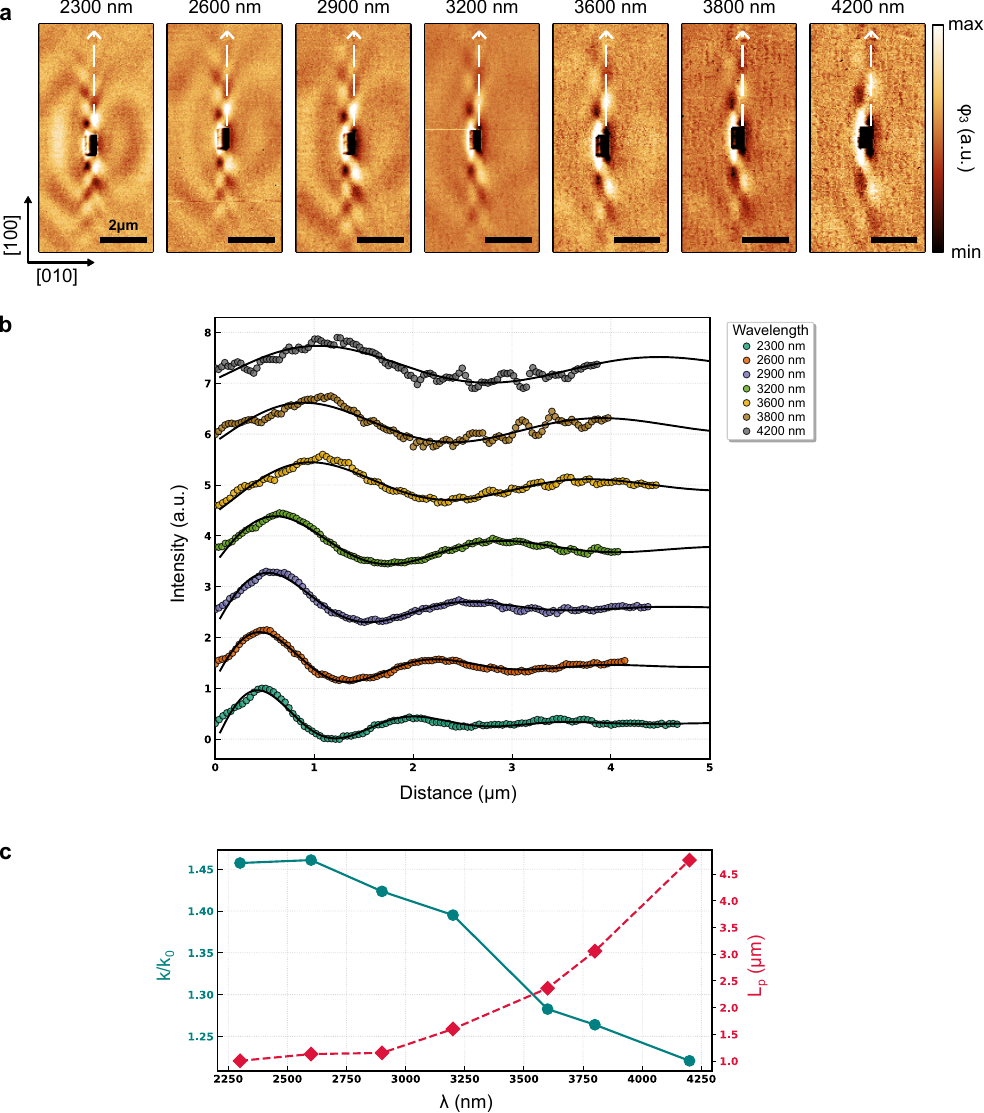}
    \caption{\textbf{Polariton propagation length on a \SI{45}{\nano \meter} flake launched by a rod antenna.} \textbf{a} sSNOM phase maps from which decay profiles are extracted along the white dashed lines. \textbf{b} Extracted profiles fitted with an exponentially decaying function. Curves are offset for clarity. \textbf{c} Retrieved wavevector and propagation length as a function of wavelength.}
    \label{fig:prop_rod}
\end{figure}

To quantify the propagation length of mid-IR polaritons in MoOCl$_2$, we extract decay profiles along the propagation direction for the measurements taken on the 45~nm flake at different wavelengths. We investigate launching from both a disk antenna (Fig.~\ref{fig:prop_disk}) and a rod antenna (Fig.~\ref{fig:prop_rod}).

In panel a of each figure, we report the near-field s-SNOM phase maps at different wavelengths. The white arrow indicates the direction along which the polariton profiles are extracted. Panel b shows the corresponding line profiles (offset for clarity), together with fits (black lines) using a decaying exponential function $f(x) = \exp(-\alpha x)\cos(kx+\phi)+C$. From these fits, we extract both the polariton wavevector $k$ and the propagation length $L_p=1/\alpha$. 

The results are summarized in panel c, where we plot the extracted wavevector normalized to the free-space value, $k/k_0$, and the propagation length as a function of wavelength. We observe a decrease in the polariton wavevector accompanied by an increase in the propagation length as the wavelength increases. This trend is expected, as higher confinement (larger $k/k_0$) is generally associated with stronger losses and thus shorter propagation distances.

The analysis is performed up to wavelengths until the canalization condition, as beyond this wavelength polaritons become strongly damped, making it difficult to reliably extract a propagation length from the experimental data. 

The increased damping observed near the canalization frequency can be attributed to the vanishing of the $\varepsilon_y$ component, which leads to enhanced electric fields in the material. As a result, even moderate intrinsic losses are effectively amplified, yielding a reduction of the propagation length in the vicinity of the canalization condition.

\begin{figure}[h!t]
    \centering
    \includegraphics[width=1.0\linewidth]{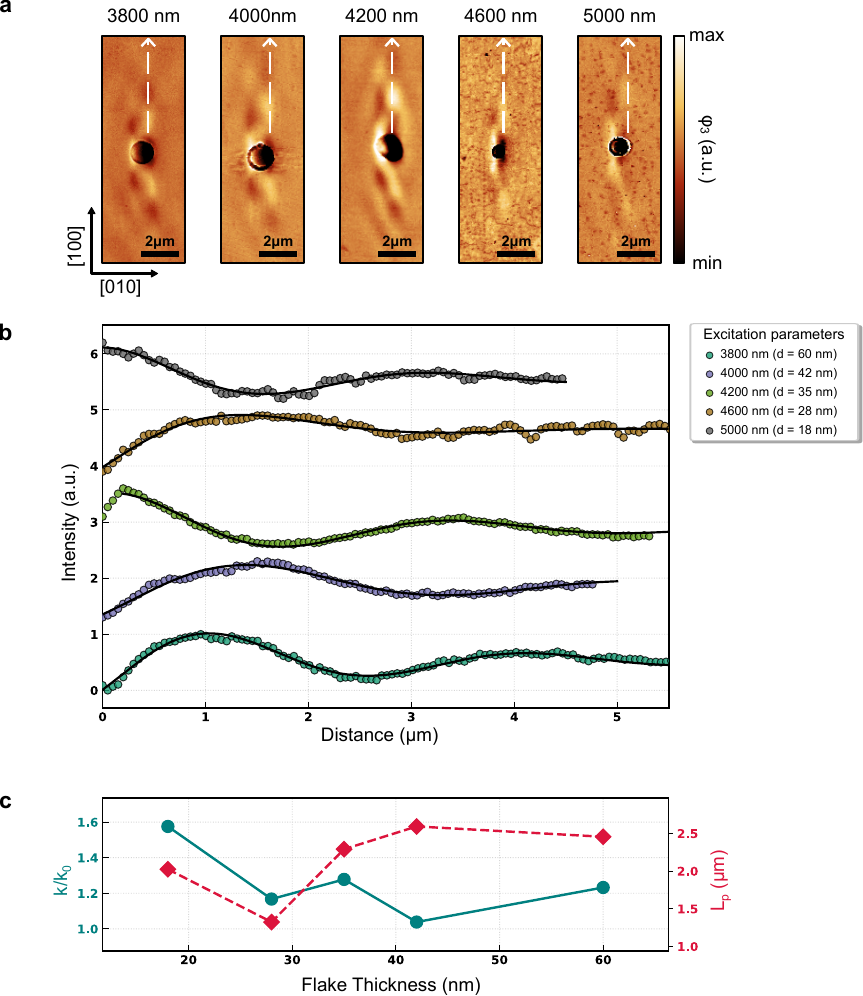}
    \caption{\textbf{Polariton propagation length at different film thicknesses.} \textbf{a} sSNOM phase maps from which decay profiles are extracted along the white dashed lines. Each map is taken at a different wavelength (just before the canalization condition) and corresponds to a different flake thickness (in order of decreasing thickness, with same values as in Fig. \ref{fig:all_thickness_fit}.). \textbf{b} Extracted profiles fitted with an exponentially decaying function. Curves are offset for clarity. \textbf{c} Retrieved wavevector and propagation length for the different flake thicknesses (and thus different canalization frequencies).}
    \label{fig:prop_thickness}
\end{figure}

To further investigate the evolution of the polariton propagation, we performed a thickness-dependent analysis of the polariton propagation length at wavelengths just below the canalization condition where it is easier to extract the propagation length (Fig.~\ref{fig:prop_thickness}). We observe that decreasing the flake thickness (thus redshifting the canalization frequency) is accompanied by an increase in the polariton momentum and a corresponding decrease in the propagation length. This behavior is consistent with the fact that achieving canalization at longer wavelengths requires thinner flakes, which results in stronger confinement of the polariton modes. As expected, increased confinement leads to enhanced losses and therefore shorter propagation distances.

\section{Increasing canalization mode confinement}\label{Canalization_confinement}

The moderate confinement of the canalized modes observed in the present geometry is mainly caused by the large negative permittivity along the [100] direction, which shifts the canalized IFCs toward relatively small in-plane momenta. To evaluate possible routes toward stronger compression, we analyze the confinement of the IFCs as a function of film thickness and wavelength. From the conic fit of the analytical IFCs, the value of the in-plane momentum at the \(k_y=0\) intercept can be extracted. Since the IFCs are nearly straight at the canalization wavelength, this value also represents the characteristic momentum of the whole canalized mode.

\begin{figure}[h!t]
    \centering
    \includegraphics[width=1\linewidth]{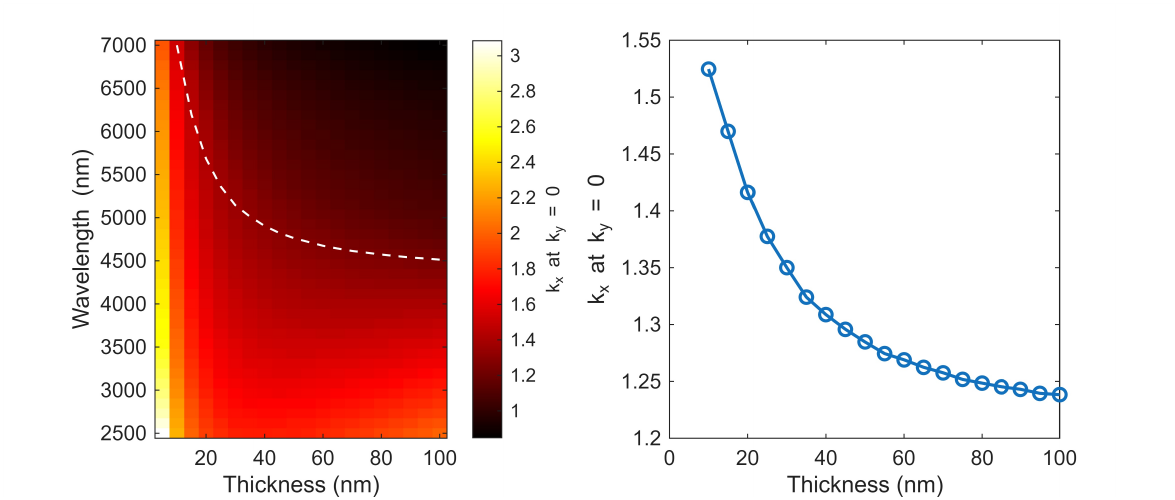}
    \caption{\textbf{Polariton confinement as a function of flake thickness.} Map of the extracted in-plane momentum \(k_x(k_y=0)\) as a function of film thickness and wavelength (left). The white dashed line indicates the canalization wavelength extracted for each thickness. Confinement factor extracted along the canalization line (right).} 
    \label{fig:confinement_thickness}
\end{figure}

Fig.~\ref{fig:confinement_thickness} shows the extracted \(k_x(k_y=0)\) as a function of film thickness and wavelength. The white dashed line marks the canalization wavelength. The corresponding confinement factor extracted along this line is also shown. The result demonstrates that thinner flakes shift the canalized response toward larger in-plane momentum, and therefore provide a direct route to stronger modal compression.

\begin{figure}[h!t]
    \centering
    \includegraphics[width=1\linewidth]{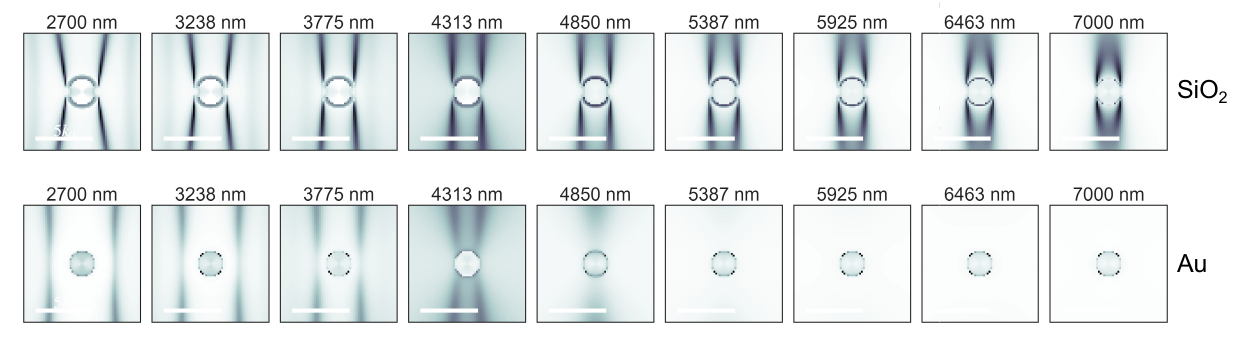}
    \caption{\textbf{Polariton dispersion on SiO$_2$ and Au substrates.} Two-dimensional transfer-matrix calculations of the polaritonic response for a MoOCl$_2$ film placed on a SiO\(_2\) substrate (top) and an Au substrate (bot).}
    \label{fig:confinement_substrate}
\end{figure}

A second route for enhanced confinemnt is substrate engineering. Fig.~\ref{fig:confinement_substrate} compares two-dimensional transfer-matrix calculations for the same film placed on a SiO\(_2\) substrate and on an Au substrate. In the SiO\(_2\)-supported geometry, the lowest-order mode dominates the response and canalization occurs at relatively small momentum. In contrast, for the Au-supported geometry, the metallic boundary modifies the modal symmetry and suppresses the first mode through symmetry and screening at the metal interface \cite{menabde2022near}. Canalization can therefore occur on a higher order branch located at larger in-plane momenta, leading to stronger confinement.

\begin{figure}[h!t]
    \centering
    \includegraphics[width=0.8\linewidth]{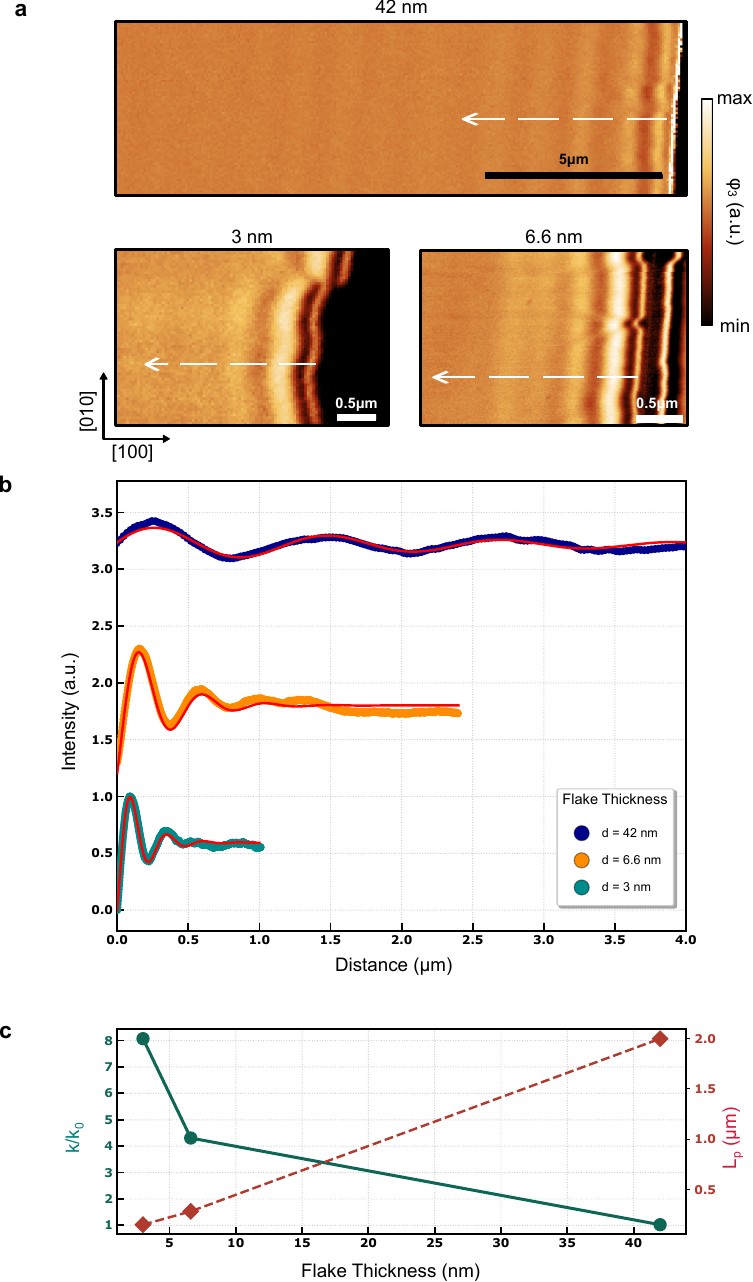}
    \caption{\textbf{Polariton propagation length at different film thicknesses.} \textbf{a} sSNOM phase maps from which decay profiles are extracted along the white dashed lines. Each map is taken at the same wavelength 2300 nm and corresponds to a different flake thickness. \textbf{b} Extracted profiles fitted with an exponentially decaying function. Curves are offset for clarity. \textbf{c} Retrieved wavevector and propagation length for the different flake thicknesses.}
    \label{fig:prop_thickness_thin_films}
\end{figure}

\begin{figure}[h!t]
    \centering
    \includegraphics[width=0.5\linewidth]{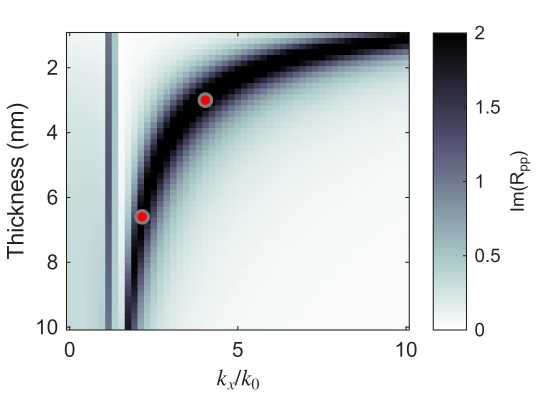}
    \caption{\textbf{Comparison of experimental momenta with transfer-matrix theory}. Experimental tip-launched momenta ($k_{exp} = 2k_{SPP}$) from the data in Fig. \ref{fig:prop_thickness_thin_films} overlayed on transfer matrix calculations for MoOCl$_2$ SPPs along the $k_x$ direction at $\lambda =$ \SI{2300}{\nano \meter}.}
    \label{fig:conf_thick_TTM}
\end{figure}

Furthermore, we performed measurements at a fixed wavelength of 2300~nm on ultrathin films with thicknesses of 3~nm and 6.6~nm to demonstrate that SPP modes can still be sustained in this extreme thickness regime. The results were compared with those obtained for a 42~nm-thick film, and the extracted polariton momentum and propagation length as functions of flake thickness are summarized in Fig.~\ref{fig:prop_thickness_thin_films}. As the film thickness decreases to 3~nm, the plasmon confinement increases dramatically, reaching values up to eight times larger than those observed in the thicker films, albeit at the expense of reduced propagation length. We also confirm (Fig. \ref{fig:conf_thick_TTM}) that the experimental momenta for the two thinner flakes are compatible with being tip-launched polaritons ($k_{exp} = 2k_{SPP}$ \cite{mancini2022near}) along the $k_x$ direction reflected by the flake edge.

\clearpage

\bibliography{sn-bibliography.bib}% common bib file

@article{herzig2024high,
  title={High-quality nanocavities through multimodal confinement of hyperbolic polaritons in hexagonal boron nitride},
  author={Herzig Sheinfux, Hanan and Orsini, Lorenzo and Jung, Minwoo and Torre, Iacopo and Ceccanti, Matteo and Marconi, Simone and Maniyara, Rinu and Barcons Ruiz, David and H{\"o}tger, Alexander and Bertini, Ricardo and others},
  journal={Nature materials},
  volume={23},
  number={4},
  pages={499--505},
  year={2024},
  publisher={Nature Publishing Group UK London}
}

@article{low2017polaritons,
  title={Polaritons in layered two-dimensional materials},
  author={Low, Tony and Chaves, Andrey and Caldwell, Joshua D and Kumar, Anshuman and Fang, Nicholas X and Avouris, Phaedon and Heinz, Tony F and Guinea, Francisco and Martin-Moreno, Luis and Koppens, Frank},
  journal={Nature materials},
  volume={16},
  number={2},
  pages={182--194},
  year={2017},
  publisher={Nature Publishing Group UK London}
}

@article{jackering2025tailoring,
  title={Tailoring hBN's Phonon Polaritons with the Plasmonic Phase-Change Material In3SbTe2},
  author={J{\"a}ckering, Lina and Moos, Aaron and Conrads, Lukas and Li, Yiheng and Rothstein, Alexander and Malik, Dominique and Watanabe, Kenji and Taniguchi, Takashi and Wuttig, Matthias and Stampfer, Christoph and others},
  journal={arXiv preprint arXiv:2504.18418},
  year={2025}
}

@article{li2015hyperbolic,
  title={Hyperbolic phonon-polaritons in boron nitride for near-field optical imaging and focusing},
  author={Li, Peining and Lewin, Martin and Kretinin, Andrey V and Caldwell, Joshua D and Novoselov, Kostya S and Taniguchi, Takashi and Watanabe, Kenji and Gaussmann, Fabian and Taubner, Thomas},
  journal={Nature communications},
  volume={6},
  number={1},
  pages={7507},
  year={2015},
  publisher={Nature Publishing Group UK London}
}

@article{Zhang2021,
  author    = {Zhang, Long and
               Wu, Fengcheng and
               Hou, Shaocong and
               Zhang, Zhe and
               Chou, Yu-Hsun and
               Watanabe, Kenji and
               Taniguchi, Takashi and
               Forrest, Stephen R. and
               Deng, Hui},
  title     = {Van der Waals heterostructure polaritons with moir{\'e}-induced nonlinearity},
  journal   = {Nature},
  year      = {2021},
  volume    = {591},
  number    = {7848},
  pages     = {61--65},
  doi       = {10.1038/s41586-021-03228-5},
  url       = {https://doi.org/10.1038/s41586-021-03228-5},
  issn      = {1476-4687}
}

@article{Basov2016,
  author    = {Basov, D. N. and
               Fogler, M. M. and
               Garc{\'\i}a de Abajo, F. J.},
  title     = {Polaritons in van der Waals materials},
  journal   = {Science},
  year      = {2016},
  volume    = {354},
  number    = {6309},
  pages     = {aag1992},
  doi       = {10.1126/science.aag1992},
  url       = {https://www.science.org/doi/abs/10.1126/science.aag1992}
}

@article{Dai2014,
  title = {Tunable Phonon Polaritons in Atomically Thin van der Waals Crystals of Boron Nitride},
  author = {Dai, Shuang and Fei, Zhe and Ma, Qiang and Rodin, Andrey S. and Wagner, Martin and McLeod, Alexander S. and Liu, Mengkun and Gannett, Wesley and Regan, William and Watanabe, Kenji and Taniguchi, Takashi and Thiemens, Mark and Dominguez, Gerardo and Castro Neto, A. H. and Zettl, Alex and Keilmann, Fritz and Jarillo-Herrero, Pablo and Fogler, M. M. and Basov, D. N.},
  journal = {Science},
  volume = {343},
  number = {6175},
  pages = {1125--1129},
  year = {2014},
  doi = {10.1126/science.1246833},
  url = {https://www.science.org/doi/abs/10.1126/science.1246833}
}

@article{Ma2018,
  author    = {Ma, Weiliang and Alonso-Gonz{\'a}lez, Pablo and Li, Shaojuan and
               Nikitin, Alexey Y. and Yuan, Jian and Mart{\'\i}n-S{\'a}nchez, Javier and
               Taboada-Guti{\'e}rrez, Javier and Amenabar, Iban and Li, Peining and
               V{\'e}lez, Sa{\"u}l and Tollan, Christopher and Dai, Zhigao and
               Zhang, Yupeng and Sriram, Sharath and Kalantar-Zadeh, Kourosh and
               Lee, Shuit-Tong and Hillenbrand, Rainer and Bao, Qiaoliang},
  title     = {In-plane anisotropic and ultra-low-loss polaritons in a natural van der Waals crystal},
  journal   = {Nature},
  volume    = {562},
  number    = {7728},
  pages     = {557--562},
  year      = {2018},
  doi       = {10.1038/s41586-018-0618-9}
}

@article{Correas-Serrano2017,
  title = {Plasmon canalization and tunneling over anisotropic metasurfaces},
  author = {Correas-Serrano, Diego and Al\`u, Andrea and Gomez-Diaz, J. Sebastian},
  journal = {Physical Review B},
  volume = {96},
  number = {7},
  pages = {075436},
  year = {2017},
  publisher = {American Physical Society},
  doi = {10.1103/PhysRevB.96.075436},
  url = {https://link.aps.org/doi/10.1103/PhysRevB.96.075436}
}

@article{kowalski2025ultraconfined,
  title={Ultraconfined terahertz phonon polaritons in hafnium dichalcogenides},
  author={Kowalski, Ryan A and Mueller, Niclas S and {\'A}lvarez-P{\'e}rez, Gonzalo and Obst, Maximilian and Diaz-Granados, Katja and Carini, Giulia and Senarath, Aditha and Dixit, Saurabh and Niemann, Richarda and Iyer, Raghunandan B and others},
  journal={Nature Materials},
  volume={24},
  number={11},
  pages={1735--1741},
  year={2025},
  publisher={Nature Publishing Group UK London}
}

@article{passler2022hyperbolic,
  title={Hyperbolic shear polaritons in low-symmetry crystals},
  author={Passler, Nikolai C and Ni, Xiang and Hu, Guangwei and Matson, Joseph R and Carini, Giulia and Wolf, Martin and Schubert, Mathias and Al{\`u}, Andrea and Caldwell, Joshua D and Folland, Thomas G and others},
  journal={Nature},
  volume={602},
  number={7898},
  pages={595--600},
  year={2022},
  publisher={Nature Publishing Group UK London}
}

@article{wang2020fermi,
  title={Fermi liquid behavior and colossal magnetoresistance in layered MoOC l 2},
  author={Wang, Zhi and Huang, Meng and Zhao, Jianzhou and Chen, Cong and Huang, Haoliang and Wang, Xiangqi and Liu, Ping and Wang, Jianlin and Xiang, Junxiang and Feng, Chao and others},
  journal={Physical Review Materials},
  volume={4},
  number={4},
  pages={041001},
  year={2020},
  publisher={APS}
}

@article{Li2020,
  author    = {Li, Peining and Hu, Guangwei and Dolado, Irene and Tymchenko, Mykhailo and Qiu, Cheng-Wei and Alfaro-Mozaz, Francisco Javier and Casanova, Fèlix and Hueso, Luis E. and Liu, Song and Edgar, James H. and Vélez, Saül and Alu, Andrea and Hillenbrand, Rainer},
  title     = {Collective near-field coupling and nonlocal phenomena in infrared-phononic metasurfaces for nano-light canalization},
  journal   = {Nature Communications},
  volume    = {11},
  number    = {1},
  pages     = {3663},
  year      = {2020},
  doi       = {10.1038/s41467-020-17425-9},
  url       = {https://doi.org/10.1038/s41467-020-17425-9}
}

@article{Hu2022,
  author    = {Hu, Hai and Chen, Na and Teng, Hanchao and Yu, Renwen and Qu, Yunpeng and Sun, Jianzhe and Xue, Mengfei and Hu, Debo and Wu, Bin and Li, Chi and Chen, Jianing and Liu, Mengkun and Sun, Zhipei and Liu, Yunqi and Li, Peining and Fan, Shanhui and Garc{\'i}a de Abajo, F. Javier and Dai, Qing},
  title     = {Doping-driven topological polaritons in graphene/$\alpha$-MoO$_3$ heterostructures},
  journal   = {Nature Nanotechnology},
  volume    = {17},
  number    = {9},
  pages     = {940--946},
  year      = {2022},
  doi       = {10.1038/s41565-022-01185-2},
  url       = {https://doi.org/10.1038/s41565-022-01185-2}
}

@article{Zhou2025,
  author    = {Zhou, Lei and Ni, Xiang and Wang, Zerui and Renzi, Enrico M. and Xu, Junbo and Zhou, Zhou and Yin, Yu and Yin, Yanzhen and Song, Renkang and Zhao, Zhichen and Yu, Ke and Huang, Di and Wang, Zhanshan and Cheng, Xinbin and Al\`u, Andrea and Jiang, Tao},
  title     = {Engineering shear polaritons in 2D twisted heterostructures},
  journal   = {Nature Communications},
  volume    = {16},
  number    = {1},
  pages     = {2953},
  year      = {2025},
  doi       = {10.1038/s41467-025-58197-4},
  url       = {https://doi.org/10.1038/s41467-025-58197-4}
}

@article{gomez2015hyperbolic,
  title={Hyperbolic plasmons and topological transitions over uniaxial metasurfaces},
  author={Gomez-Diaz, J Sebastian and Tymchenko, Mykhailo and Alu, Andrea},
  journal={Physical review letters},
  volume={114},
  number={23},
  pages={233901},
  year={2015},
  publisher={APS}
}

@article{zhang2025ultimate,
  title={Ultimate tuning of hyperbolic phonon polaritons},
  author={Zhang, Linglong and Ding, Xiaojunjie and Dong, Jiahao and Fan, Jiang and Wu, Wei and Ren, Mengxin and Luo, Weiwei and Cai, Wei and Xu, Jingjun},
  journal={Science Advances},
  volume={11},
  number={50},
  pages={eadz6278},
  year={2025},
  publisher={American Association for the Advancement of Science}
}

@article{ou2025natural,
  title={Natural van der Waals Canalization Lens for Non-Destructive Nanoelectronic Circuit Imaging and Inspection},
  author={Ou, Qingdong and Xue, Shuwen and Ma, Weiliang and Yang, Jiong and Si, Guangyuan and Liu, Lu and Zhong, Gang and Liu, Jingying and Xie, Zongyuan and Xiao, Ying and others},
  journal={Advanced Materials},
  pages={2504526},
  year={2025},
  publisher={Wiley Online Library}
}

@article{Terán-García,
  author  = {Terán-García, Enrique and Lanza, Christian and Voronin, Kirill and Martín-Sánchez, Javier and Nikitin, Alexey Y. and Tarazaga Martín-Luengo, Aitana and Alonso-González, Pablo},
  title   = {Real-Space Visualization of Canalized Ray Polaritons in a Single Van der Waals Thin Slab},
  journal = {Nano Letters},
  year    = {2025},
  volume  = {25},
  number  = {6},
  pages   = {2203--2209},
  doi     = {10.1021/acs.nanolett.4c05277},
  url     = {https://doi.org/10.1021/acs.nanolett.4c05277}
}

@article{Wang,
  author  = {Wang, Kun and Huang, Zhongao and Xiong, Langlang and Wang, Kai and Bai, Yihua and Long, Hua and Deng, Nan and Wang, Bing and Hu, Guangwei and Lu, Peixiang},
  title   = {Observation of canalized phonon polaritons in a single $\alpha$-MoO$_3$ flake},
  journal = {Optica},
  year    = {2025},
  volume  = {12},
  number  = {3},
  pages   = {343--349},
  doi     = {10.1364/OPTICA.547698},
  url     = {https://opg.optica.org/optica/abstract.cfm?URI=optica-12-3-343}
}

@article{TresguerresMata2024,
  author  = {Tresguerres-Mata, Ana I. F. and Lanza, Christian and Taboada-Gutiérrez, Javier and Matson, Joseph R. and Álvarez-Pérez, Gonzalo and Isobe, Masahiko and Tarazaga Martín-Luengo, Aitana and Duan, Jiahua and Partel, Stefan and Vélez, María and Martín-Sánchez, Javier and Nikitin, Alexey Y. and Caldwell, Joshua D. and Alonso-González, Pablo},
  title   = {Observation of naturally canalized phonon polaritons in LiV$_2$O$_5$ thin layers},
  journal = {Nature Communications},
  year    = {2024},
  volume  = {15},
  number  = {1},
  pages   = {2696},
  doi     = {10.1038/s41467-024-46935-z},
  url     = {https://doi.org/10.1038/s41467-024-46935-z}
}

@article{Díaz-Núñez,
  author  = {Díaz-Núñez, Pablo and Lanza, Christian and Wang, Ziwei and Kravets, Vasyl G. and Duan, Jiahua and Álvarez-Cuervo, José and Tarazaga Martín-Luengo, Aitana and Grigorenko, Alexander N. and Yang, Qian and Paarmann, Alexander and Caldwell, Joshua and Alonso-González, Pablo and Mishchenko, Artem},
  title   = {Visualization of topological shear polaritons in gypsum thin films},
  journal = {Science Advances},
  year    = {2025},
  volume  = {11},
  number  = {29},
  pages   = {eadw3452},
  doi     = {10.1126/sciadv.adw3452},
  url     = {https://www.science.org/doi/abs/10.1126/sciadv.adw3452}
}

@article{Xing2024,
  author  = {Xing, Qiaoxia and Zhang, Jiasheng and Fang, Yuqiang and Song, Chaoyu and Zhao, Tuoyu and Mou, Yanlin and Wang, Chong and Ma, Junwei and Xie, Yuangang and Huang, Shenyang and Mu, Lei and Lei, Yuchen and Shi, Wu and Huang, Fuqiang and Yan, Hugen},
  title   = {Tunable anisotropic van der Waals films of 2M-WS2 for plasmon canalization},
  journal = {Nature Communications},
  year    = {2024},
  volume  = {15},
  pages   = {2623},
  doi     = {10.1038/s41467-024-46963-9},
  url     = {https://doi.org/10.1038/s41467-024-46963-9}
}

@article{Venturi2024,
  author  = {Venturi, Giacomo and Mancini, Andrea and Melchioni, Nicola and Chiodini, Stefano and Ambrosio, Antonio},
  title   = {Visible-frequency hyperbolic plasmon polaritons in a natural van der Waals crystal},
  journal = {Nature Communications},
  year    = {2024},
  volume  = {15},
  number  = {1},
  pages   = {9727},
  doi     = {10.1038/s41467-024-53988-7},
  url     = {https://doi.org/10.1038/s41467-024-53988-7}
}

@article{zhang2021interface,
  title={Interface nano-optics with van der Waals polaritons},
  author={Zhang, Qing and Hu, Guangwei and Ma, Weiliang and Li, Peining and Krasnok, Alex and Hillenbrand, Rainer and Al{\`u}, Andrea and Qiu, Cheng-Wei},
  journal={Nature},
  volume={597},
  number={7875},
  pages={187--195},
  year={2021},
  publisher={Nature Publishing Group UK London}
}

@article{he2022anisotropy,
  title={Anisotropy and modal hybridization in infrared nanophotonics using low-symmetry materials},
  author={He, Mingze and Folland, Thomas G and Duan, Jiahua and Alonso-Gonz{\'a}lez, Pablo and De Liberato, Simone and Paarmann, Alexander and Caldwell, Joshua D},
  journal={Acs photonics},
  volume={9},
  number={4},
  pages={1078--1095},
  year={2022},
  publisher={ACS Publications}
}

@article{alvarez2022negative,
  title={Negative reflection of nanoscale-confined polaritons in a low-loss natural medium},
  author={{\'A}lvarez-P{\'e}rez, Gonzalo and Duan, Jiahua and Taboada-Guti{\'e}rrez, Javier and Ou, Qingdong and Nikulina, Elizaveta and Liu, Song and Edgar, James H and Bao, Qiaoliang and Giannini, Vincenzo and Hillenbrand, Rainer and others},
  journal={Science advances},
  volume={8},
  number={29},
  pages={eabp8486},
  year={2022},
  publisher={American Association for the Advancement of Science}
}

@article{ocelic2006pseudoheterodyne,
  title={Pseudoheterodyne detection for background-free near-field spectroscopy},
  author={Ocelic, Nenad and Huber, Andreas and Hillenbrand, Rainer},
  journal={Applied Physics Letters},
  volume={89},
  number={10},
  year={2006},
  publisher={AIP Publishing}
}

@article{xu2026programmable,
  title={Programmable polariton canalization in reconfigurable metasurfaces},
  author={Xu, Yang and Su, Xiaoqiang and Deng, Fusheng and Wang, Yuqian and Yang, Yaping and Alonso-Gonz{\'a}lez, Pablo and Chen, Hong and Li, Jiafang and Duan, Jiahua and Guo, Zhiwei},
  journal={Science Advances},
  volume={12},
  number={2},
  pages={eaea0072},
  year={2026},
  publisher={American Association for the Advancement of Science}
}

@article{zhang2025manipulating,
  title={Manipulating hyperbolic plasmon polaritons at near-infrared in an anisotropic van der Waals crystal},
  author={Zhang, Yuxin and Li, Yaolong and Xiao, Jingying and Luo, Yijie and Li, Xiaofang and Tang, Jinglin and Xu, Xiayuan and Jiang, Pengzuo and Zhang, Guanyu and Tang, Huilin and others},
  journal={Nano Letters},
  volume={25},
  number={43},
  pages={15534--15541},
  year={2025},
  publisher={ACS Publications}
}

@article{chang2022field,
  title={Field canalization using anisotropic 2D plasmonics},
  author={Chang, Po-Han and Lin, Charles and Helmy, Amr S},
  journal={npj 2D Materials and Applications},
  volume={6},
  number={1},
  pages={5},
  year={2022},
  publisher={Nature Publishing Group UK London}
}

@article{zhang2025phonon,
  title={Phonon engineering enables hyperbolic asymptotic line polaritons},
  author={Zhang, Shu and Ma, Puyi and You, Oubo and Zhou, Shenghan and Feng, Kaijun and Yuan, Hongyi and Zhang, Jinhao and Wu, Chenchen and Luo, Yang and Yang, Bei and others},
  journal={Nature Nanotechnology},
  pages={1--6},
  year={2025},
  publisher={Nature Publishing Group UK London}
}

@article{li2025broadband,
  title={Broadband near-infrared hyperbolic polaritons in MoOCl2},
  author={Li, Yaolong and Zhang, Yuxin and Zhang, Weizhe and Li, Xiaofang and Tang, Jinglin and Xiao, Jingying and Zhang, Guanyu and Liao, Xin and Jiang, Pengzuo and Liu, Qinyun and others},
  journal={Nature Communications},
  volume={16},
  number={1},
  pages={6172},
  year={2025},
  publisher={Nature Publishing Group UK London}
}

@article{ruta2025good,
  title={Good plasmons in a bad metal},
  author={Ruta, Francesco L and Shao, Yinming and Acharya, Swagata and Mu, Anqi and Jo, Na Hyun and Ryu, Sae Hee and Balatsky, Daria and Su, Yifan and Pashov, Dimitar and Kim, Brian SY and others},
  journal={Science},
  volume={387},
  number={6735},
  pages={786--791},
  year={2025},
  publisher={American Association for the Advancement of Science}
}

@article{sternbach2023negative,
  title={Negative refraction in hyperbolic hetero-bicrystals},
  author={Sternbach, AJ and Moore, SL and Rikhter, Andrey and Zhang, Shuai and Jing, Ran and Shao, Yinming and Kim, BSY and Xu, Suheng and Liu, Song and Edgar, JH and others},
  journal={Science},
  volume={379},
  number={6632},
  pages={555--557},
  year={2023},
  publisher={American Association for the Advancement of Science}
}

@article{hu2023gate,
  title={Gate-tunable negative refraction of mid-infrared polaritons},
  author={Hu, Hai and Chen, Na and Teng, Hanchao and Yu, Renwen and Xue, Mengfei and Chen, Ke and Xiao, Yuchuan and Qu, Yunpeng and Hu, Debo and Chen, Jianing and others},
  journal={Science},
  volume={379},
  number={6632},
  pages={558--561},
  year={2023},
  publisher={American Association for the Advancement of Science}
}

@article{teng2024steering,
  title={Steering and cloaking of hyperbolic polaritons at deep-subwavelength scales},
  author={Teng, Hanchao and Chen, Na and Hu, Hai and Garc{\'\i}a de Abajo, F Javier and Dai, Qing},
  journal={Nature Communications},
  volume={15},
  number={1},
  pages={4463},
  year={2024},
  publisher={Nature Publishing Group UK London}
}

@article{datta2022highly,
  title={Highly nonlinear dipolar exciton-polaritons in bilayer MoS2},
  author={Datta, Biswajit and Khatoniar, Mandeep and Deshmukh, Prathmesh and Thouin, F{\'e}lix and Bushati, Rezlind and De Liberato, Simone and Cohen, Stephane Kena and Menon, Vinod M},
  journal={Nature communications},
  volume={13},
  number={1},
  pages={6341},
  year={2022},
  publisher={Nature Publishing Group UK London}
}

@article{hu2020phonon,
  title={Phonon polaritons and hyperbolic response in van der Waals materials},
  author={Hu, Guangwei and Shen, Jialiang and Qiu, Cheng-Wei and Al{\`u}, Andrea and Dai, Siyuan},
  journal={Advanced Optical Materials},
  volume={8},
  number={5},
  pages={1901393},
  year={2020},
  publisher={Wiley Online Library}
}

@article{fei2012gate,
  title={Gate-tuning of graphene plasmons revealed by infrared nano-imaging},
  author={Fei, Zhe and Rodin, AS and Andreev, Gregory O and Bao, Wenzhong and McLeod, AS and Wagner, M and Zhang, LM and Zhao, Zeng and Thiemens, M and Dominguez, Gerardo and others},
  journal={Nature},
  volume={487},
  number={7405},
  pages={82--85},
  year={2012},
  publisher={Nature Publishing Group UK London}
}

@article{chen2012optical,
  title={Optical nano-imaging of gate-tunable graphene plasmons},
  author={Chen, Jianing and Badioli, Michela and Alonso-Gonz{\'a}lez, Pablo and Thongrattanasiri, Sukosin and Huth, Florian and Osmond, Johann and Spasenovi{\'c}, Marko and Centeno, Alba and Pesquera, Amaia and Godignon, Philippe and others},
  journal={Nature},
  volume={487},
  number={7405},
  pages={77--81},
  year={2012},
  publisher={Nature Publishing Group UK London}
}

@article{alcaraz2018probing,
  title={Probing the ultimate plasmon confinement limits with a van der Waals heterostructure},
  author={Alcaraz Iranzo, David and Nanot, S{\'e}bastien and Dias, Eduardo JC and Epstein, Itai and Peng, Cheng and Efetov, Dmitri K and Lundeberg, Mark B and Parret, Romain and Osmond, Johann and Hong, Jin-Yong and others},
  journal={Science},
  volume={360},
  number={6386},
  pages={291--295},
  year={2018},
  publisher={American Association for the Advancement of Science}
}

@article{ciraci2012probing,
  title={Probing the ultimate limits of plasmonic enhancement},
  author={Cirac{\`\i}, Cristian and Hill, RT and Mock, JJ and Urzhumov, Yaroslav and Fern{\'a}ndez-Dom{\'\i}nguez, AI and Maier, SA and Pendry, JB and Chilkoti, Ashutosh and Smith, DR},
  journal={Science},
  volume={337},
  number={6098},
  pages={1072--1074},
  year={2012},
  publisher={American Association for the Advancement of Science}
}

@book{dressel2002electrodynamics,
  title={Electrodynamics of solids: optical properties of electrons in matter},
  author={Dressel, Martin and Gr{\"u}ner, George},
  year={2002},
  publisher={Cambridge university press}
}

@article{Melchioni2025,
  author  = {Melchioni, Nicola and Mancini, Andrea and Nan, Lin and Efimova, Anastasiia and Venturi, Giacomo and Ambrosio, Antonio},
  title   = {Giant Optical Anisotropy in a Natural van der Waals Hyperbolic Crystal for Visible Light Low-Loss Polarization Control},
  journal = {ACS Nano},
  year    = {2025},
  volume  = {19},
  number  = {27},
  pages   = {25413--25421},
  doi     = {10.1021/acsnano.5c07323},
  url     = {https://doi.org/10.1021/acsnano.5c07323}
}

@article{zhao2020highly,
  title={Highly anisotropic two-dimensional metal in monolayer MoOCl 2},
  author={Zhao, Jianzhou and Wu, Weikang and Zhu, Jiaojiao and Lu, Yunhao and Xiang, Bin and Yang, Shengyuan A},
  journal={Physical Review B},
  volume={102},
  number={24},
  pages={245419},
  year={2020},
  publisher={APS}
}

@article{Liu2025,
  author       = {Liu, Lu and Xiong, Langlang and Wang, Chongwu and Bai, Yihua and Ma, Weiliang and Wang, Yupeng and Li, Peining and Li, Guogang and Wang, Qi Jie and Garcia-Vidal, Francisco J. and Dai, Zhigao and Hu, Guangwei},
  title        = {Long-range hyperbolic polaritons on a non-hyperbolic crystal surface},
  journal      = {Nature},
  year         = {2025},
  volume       = {644},
  number       = {8075},
  pages        = {76--82},
  doi          = {10.1038/s41586-025-09288-1},
  url          = {https://doi.org/10.1038/s41586-025-09288-1}
}

@article{ghosh2026spatiotemporal,
  title={Spatiotemporal visualization of long-range anisotropic plasmon polaritons in hyperbolic MoOCl2},
  author={Ghosh, Atreyie and Raab, Calvin and Spellberg, Joseph L and Mohan, Aishani and Munawar, Muneeza and Rieger, Janek and King, Sarah B},
  journal={Nature Communications},
  year={2026},
  publisher={Nature Publishing Group UK London}
}

@article{melchioni2026anisotropic,
  title={Anisotropic electron gas in a hyperbolic van der Waals material},
  author={Melchioni, Nicola and Mancini, Andrea and Ambrosio, Antonio},
  journal={arXiv preprint arXiv:2602.01072},
  year={2026}
}

@article{ermolaev2026giant,
  title={Giant optical anisotropy and visible-frequency epsilon-near-zero in hyperbolic van der Waals MoOCl2},
  author={Ermolaev, Georgy and Toksumakov, Adilet and Slavich, Aleksandr and Minnekhanov, Anton and Tselikov, Gleb and Mazitov, Arslan and Kruglov, Ivan and Tikhonowski, Gleb and Mironov, Mikhail and Radko, Ilya P and others},
  journal={Nano Letters},
  volume={26},
  number={13},
  pages={4329--4338},
  year={2026},
  publisher={ACS Publications}
}

@article{duan2021enabling,
  title={Enabling propagation of anisotropic polaritons along forbidden directions via a topological transition},
  author={Duan, Jiahua and {\'A}lvarez-P{\'e}rez, Gonzalo and Voronin, Kirill V and Prieto, Iv{\'a}n and Taboada-Guti{\'e}rrez, Javier and Volkov, Valentyn S and Mart{\'\i}n-S{\'a}nchez, Javier and Nikitin, Alexey Y and Alonso-Gonz{\'a}lez, Pablo},
  journal={Science advances},
  volume={7},
  number={14},
  pages={eabf2690},
  year={2021},
  publisher={American Association for the Advancement of Science}
}

@article{menabde2022near,
  title={Near-field probing of image phonon-polaritons in hexagonal boron nitride on gold crystals},
  author={Menabde, Sergey G and Boroviks, Sergejs and Ahn, Jongtae and Heiden, Jacob T and Watanabe, Kenji and Taniguchi, Takashi and Low, Tony and Hwang, Do Kyung and Mortensen, N Asger and Jang, Min Seok},
  journal={Science advances},
  volume={8},
  number={28},
  pages={eabn0627},
  year={2022},
  publisher={American Association for the Advancement of Science}
}

@book{novotny2012principles,
  title={Principles of nano-optics},
  author={Novotny, Lukas and Hecht, Bert},
  year={2012},
  publisher={Cambridge university press},
  address   = {Cambridge}
}

@article{ambrosio2018selective,
  title={Selective excitation and imaging of ultraslow phonon polaritons in thin hexagonal boron nitride crystals},
  author={Ambrosio, Antonio and Tamagnone, Michele and Chaudhary, Kundan and Jauregui, Luis A and Kim, Philip and Wilson, William L and Capasso, Federico},
  journal={Light: Science \& Applications},
  volume={7},
  number={1},
  pages={27},
  year={2018},
  publisher={Nature Publishing Group UK London}
}

@article{hu2023source,
  title={Source-configured symmetry-broken hyperbolic polaritons},
  author={Hu, Caixing and Sun, Tian and Zeng, Ying and Ma, Weiliang and Dai, Zhigao and Yang, Xiaosheng and Zhang, Xinliang and Li, Peining},
  journal={Elight},
  volume={3},
  number={1},
  pages={14},
  year={2023},
  publisher={Springer}
}

@article{zheng2024hyperbolic,
  title={Hyperbolic-to-hyperbolic transition at exceptional Reststrahlen point in rare-earth oxyorthosilicates},
  author={Zheng, Chunqi and Hu, Guangwei and Wei, Jingxuan and Ma, Xuezhi and Li, Zhipeng and Chen, Yinzhu and Ni, Zhenhua and Li, Peining and Wang, Qian and Qiu, Cheng-Wei},
  journal={Nature Communications},
  volume={15},
  number={1},
  pages={7047},
  year={2024},
  publisher={Nature Publishing Group UK London}
}

@article{shiravi2026tunable,
  title={Tunable Polariton Canalization in Natural van der Waals Oxide},
  author={Shiravi, H and Zheng, W and Rhodes, DA and Balicas, L and Zhou, HD and Ni, GX},
  journal={arXiv preprint arXiv:2604.12174},
  year={2026}
}

@article{zhou2026fundamental,
  title={Fundamental optical phenomena of strongly anisotropic polaritons at the nanoscale},
  author={Zhou, Yixi and Guo, Zhiwei and Tarazaga Mart{\'\i}n-Luengo, Aitana and Lanza, Christian and {\'A}lvarez-P{\'e}rez, Gonzalo and Yu, Chengguang and Li, Chongrui and Xia, Weixiang and {\'A}lvarez Cuervo, Jos{\'e} and Duan, Xiaoyang and others},
  journal={Nature nanotechnology},
  volume={21},
  number={1},
  pages={23--38},
  year={2026},
  publisher={Nature Publishing Group UK London}
}

@article{mancini2022near,
  title={Near-Field Retrieval of the Surface Phonon Polariton Dispersion in Free-Standing Silicon Carbide Thin Films},
  author={Mancini, Andrea and Nan, Lin and Wendisch, Fedja J and Bert{\'e}, Rodrigo and Ren, Haoran and Cort{\'e}s, Emiliano and Maier, Stefan A},
  journal={ACS Photonics},
  volume={9},
  number={11},
  pages={3696--3704},
  year={2022},
  publisher={ACS Publications}
}

@article{kaltenecker2020mono,
  title={Mono-crystalline gold platelets: a high-quality platform for surface plasmon polaritons},
  author={Kaltenecker, Korbinian J and Krauss, Enno and Casses, Laura and Geisler, Mathias and Hecht, Bert and Mortensen, N Asger and Jepsen, Peter Uhd and Stenger, Nicolas},
  journal={Nanophotonics},
  volume={9},
  number={2},
  pages={509--522},
  year={2020},
  publisher={De Gruyter}
}

@article{duan2023multiple,
  title={Multiple and spectrally robust photonic magic angles in reconfigurable $\alpha$-MoO3 trilayers},
  author={Duan, Jiahua and {\'A}lvarez-P{\'e}rez, Gonzalo and Lanza, Christian and Voronin, K and Tresguerres-Mata, Ana IF and Capote-Robayna, Nathaniel and {\'A}lvarez-Cuervo, Jos{\'e} and Tarazaga Mart{\'\i}n-Luengo, Aitana and Mart{\'\i}n-S{\'a}nchez, Javier and Volkov, Valentyn S and others},
  journal={Nature Materials},
  volume={22},
  number={7},
  pages={867--872},
  year={2023},
  publisher={Nature Publishing Group UK London}
}

@article{passler2017generalized,
  title={Generalized 4$\times$ 4 matrix formalism for light propagation in anisotropic stratified media: study of surface phonon polaritons in polar dielectric heterostructures},
  author={Passler, Nikolai Christian and Paarmann, Alexander},
  journal={Journal of the Optical Society of America B},
  volume={34},
  number={10},
  pages={2128--2139},
  year={2017},
  publisher={Optical Society of America}
}

@article{alvarez2019analytical,
  title={Analytical approximations for the dispersion of electromagnetic modes in slabs of biaxial crystals},
  author={{\'A}lvarez-P{\'e}rez, Gonzalo and Voronin, Kirill V and Volkov, Valentyn S and Alonso-Gonz{\'a}lez, Pablo and Nikitin, Alexey Y},
  journal={Physical Review B},
  volume={100},
  number={23},
  pages={235408},
  year={2019},
  publisher={APS}
}

@article{ruta2024good,
  title={Good plasmons in a bad metal},
  author={Ruta, Francesco L and Shao, Yinming and Acharya, Swagata and Mu, Anqi and Jo, Na Hyun and Ryu, Sae Hee and Balatsky, Daria and Su, Yifan and Pashov, Dimitar and Kim, Brian SY and others},
  journal={Science},
  volume={387},
  number={6735},
  pages={786--791},
  year={2025},
  publisher={American Association for the Advancement of Science}
}

@article{zheng2019mid,
  title={A mid-infrared biaxial hyperbolic van der Waals crystal},
  author={Zheng, Zebo and Xu, Ningsheng and Oscurato, Stefano L and Tamagnone, Michele and Sun, Fengsheng and Jiang, Yinzhu and Ke, Yanlin and Chen, Jianing and Huang, Wuchao and Wilson, William L and others},
  journal={Science advances},
  volume={5},
  number={5},
  pages={eaav8690},
  year={2019},
  publisher={American Association for the Advancement of Science}
}

@article{duan2025canalization,
  title={Canalization-based super-resolution imaging using an individual van der Waals thin layer},
  author={Duan, Jiahua and Mart{\'\i}n-Luengo, Aitana Tarazaga and Lanza, Christian and Partel, Stefan and Voronin, Kirill and Tresguerres-Mata, Ana Isabel F and {\'A}lvarez-P{\'e}rez, Gonzalo and Nikitin, Alexey Y and Mart{\'\i}n-S{\'a}nchez, Javier and Alonso-Gonz{\'a}lez, Pablo},
  journal={Science advances},
  volume={11},
  number={7},
  pages={eads0569},
  year={2025},
  publisher={American Association for the Advancement of Science}
}

@article{obst2023terahertz,
  title={Terahertz twistoptics--engineering canalized phonon polaritons},
  author={Obst, Maximilian and Nooerenberg, Tobias and {\'A}lvarez-P{\'e}rez, Gonzalo and de Oliveira, Thales VAG and Taboada-Guti{\'e}rrez, Javier and Feres, Fl{\'a}vio H and Kaps, Felix G and Hatem, Osama and Luferau, Andrei and Nikitin, Alexey Y and others},
  journal={ACS nano},
  volume={17},
  number={19},
  pages={19313--19322},
  year={2023},
  publisher={ACS Publications}
}

@article{chen2020configurable,
  title={Configurable phonon polaritons in twisted $\alpha$-MoO3},
  author={Chen, Mingyuan and Lin, Xiao and Dinh, Thao H and Zheng, Zhiren and Shen, Jialiang and Ma, Qiong and Chen, Hongsheng and Jarillo-Herrero, Pablo and Dai, Siyuan},
  journal={Nature materials},
  volume={19},
  number={12},
  pages={1307--1311},
  year={2020},
  publisher={Nature Publishing Group UK London}
}

@article{duan2020twisted,
  title={Twisted nano-optics: manipulating light at the nanoscale with twisted phonon polaritonic slabs},
  author={Duan, Jiahua and Capote-Robayna, Nathaniel and Taboada-Guti{\'e}rrez, Javier and {\'A}lvarez-P{\'e}rez, Gonzalo and Prieto, Iv{\'a}n and Mart{\'\i}n-S{\'a}nchez, Javier and Nikitin, Alexey Y and Alonso-Gonz{\'a}lez, Pablo},
  journal={Nano Letters},
  volume={20},
  number={7},
  pages={5323--5329},
  year={2020},
  publisher={ACS Publications}
}

@article{zhu2025multiple,
  title={Multiple hyperbolic dispersion branches and broadband canalization in a phonon-polaritonic heterostructure},
  author={Zhu, Jiaqi and Gong, Youning and Liang, Jun and Zhao, Yanyu and Cui, Zhe and Li, Delong and Ou, Qingdong and Zhang, Yupeng and Wang, Guo Ping},
  journal={Nano Letters},
  volume={25},
  number={7},
  pages={2610--2617},
  year={2025},
  publisher={ACS Publications}
}

@article{hu2020topological,
  title={Topological polaritons and photonic magic angles in twisted $\alpha$-MoO3 bilayers},
  author={Hu, Guangwei and Ou, Qingdong and Si, Guangyuan and Wu, Yingjie and Wu, Jing and Dai, Zhigao and Krasnok, Alex and Mazor, Yarden and Zhang, Qing and Bao, Qiaoliang and others},
  journal={Nature},
  volume={582},
  number={7811},
  pages={209--213},
  year={2020},
  publisher={Nature Publishing Group UK London}
}

\end{document}